\documentclass[%
reprint,
amsmath,amssymb,
aps,
]{revtex4-2}
\usepackage{xcolor}
\usepackage{graphicx}
\usepackage{dcolumn}
\graphicspath{{./graphics/}}
\usepackage{enumerate}
\usepackage{subcaption}
\usepackage{bm}

\begin{document}

\title{Motion of microswimmers in cylindrical microchannels} 

\author{Florian A. Overberg, Gerhard Gompper, and Dmitry A. Fedosov}
\affiliation{Theoretical Physics of Living Matter, Institute of Biological Information Processing and Institute for Advanced Simulation,
Forschungszentrum J\"ulich, 52425 J\"ulich, Germany \\
Email: f.overberg@fz-juelich.de, g.gompper@fz-juelich.de, d.fedosov@fz-juelich.de}

\date{\today}

\begin{abstract}
Biological and artificial microswimmers often have to propel through a variety of environments, ranging from heterogeneous suspending media to strong
geometrical confinement. Under confinement, local flow fields generated by microswimmers, and steric and hydrodynamic interactions with their environment 
determine the locomotion. We propose a squirmer-like model to describe the motion of microswimmers in cylindrical microchannels, where propulsion 
is generated by a fixed surface slip velocity.  The model is studied analytically for cylindrical swimmer shapes, and by numerical hydrodynamics simulations 
for spherical and spheroidal shapes. For the numerical simulations, we employ the dissipative particle dynamics method for modelling fluid flow.
Both the analytical model and simulations show that the propulsion force increases with increasing confinement. 
However, the swimming velocity under confinement remains lower than the swimmer speed without confinement for all investigated conditions. In simulations, 
different swimming modes (i.e. pusher, neutral, puller) are investigated, and found to play a significant role in the generation of propulsion force when a 
swimmer approaches a dead end of a capillary. Propulsion generation in confined systems is local, such that the generated flow field generally vanishes 
beyond the characteristic size of the swimmer. These results contribute to a better understanding of microswimmer force generation and propulsion under 
strong confinement, including the motion in porous media and in narrow channels.
\end{abstract}


\maketitle


\section{Introduction}

In their natural habitat, biological microswimmers (e.g. \textit{E. coli}, \textit{Paramecium}) are subject to a multitude of 
different environments, ranging from unconfined liquid media to porous-like surroundings within the soil or a biofilm 
\cite{Elgeti_PMS_2015,Bechinger_APC_2016}. Properties of these media can be very diverse, and include fluid viscosity varying over 
several orders of magnitude \cite{Kaiser_ETM_1975,Nsamela_EVM_2021}, viscoelasticity in complex fluidic environments 
\cite{Tung_FVE_2017,Krueger_BBT_2018}, and various degrees of hard and soft confinements 
\cite{Bhattacharjee_BHT_2019,Takatori_ACF_2020,Vutukuri_APV_2020}. In such environments, microswimmers are subject to intricate
steric and hydrodynamic interactions \cite{Elgeti_MSS_2016}, which they have to cope with and be able to employ for their efficient 
locomotion and navigation. Furthermore, understanding microswimmer propulsion in complex environments is relevant 
for the development of microscopic artificial motile systems capable of performing specific tasks \cite{Palagi_BIR_2018,Medina_SMR_2018}.

Most studies of locomotion under confinement have focused on microswimmer-wall 
interactions \cite{Elgeti_EMW_2013,Bechinger_APC_2016}. For example, microswimmers are subject to wall accumulation, which 
is due to their slow re-orientation (e.g. governed by rotational diffusion) after they hit a wall as well as hydrodynamic interactions\cite{Wensink_ASP_2008,Berke_HAS_2008,Elgeti_WAS_2013}.
The accumulation is even more enhanced in places with a non-zero wall curvature \cite{Fily_DSP_2014,Fily_DDD_2015,Iyer_MIPS_2023}, and this 
effect can lead to a persistent microswimmer motion along the surface \cite{Vladescu_DMB_2014,Rode_SMC_2019}. Studies of microswimmer propulsion under strong confinements are still rather scarce due to difficulties in experimental 
observations or numerical modeling. For instance, \textit{E. coli} in porous media exhibits hopping and trapping behavior 
\cite{Bhattacharjee_BHT_2019}, in contrast to the run-and-tumble motion within unconfined fluidic environments \cite{Turner_RTF_2000}.
Furthermore, trypanosomes (parasites causing sleeping sickness) can survive in and navigate through very different environments, 
including blood and several solid-like tissues \cite{Krueger_BBT_2018,Bargul_SSA_2016}. An investigation of trypanosome locomotion 
through a maze of obstacles within a microfluidic device suggested that these parasites can sometimes move more efficiently through
crowded environments \cite{Heddergott_TMR_2012}. Nevertheless, the majority of experimental and theoretical studies report 
a reduction in microswimmer speed under confinement in comparison with that within unconfined fluid conditions
\cite{Bhattacharjee_BHT_2019,Zhu_LRN_2013,Liu_PHF_2014,Creppy_EMT_2019}.

Another interesting example is the motion of microswimmers under soft deformable confinements represented by a lipid membrane and vesicles
\cite{Takatori_ACF_2020,Vutukuri_APV_2020,Le_Nagard_EBV_2022,Iyer_SAV_2022}. This active system shows a variety of dynamic
non-equilibrium vesicle shapes, including prolate geometries and structures with multiple tether-like protrusions generated
by the encapsulated swimmers. Within the tethers, the swimmers are tightly wrapped by the membrane; however, they are still able to 
propel and exert forces on the membrane to further extend the tethers. \textit{E. coli} bacteria were found not only to pull relatively long tethers, but also to transport the vesicle with a non-zero velocity despite the tight wrapping by the vesicle membrane\cite{Le_Nagard_EBV_2022}.

The examples above raise a number of scientific questions related to microswimmer propulsion under confinement. How can 
microswimmers propel and navigate through very confined environments? Do they employ the physical mechanisms and strategies 
similar to those when propelling under unconfined conditions? Does the local flow field generated by microswimmers play 
an essential role in their propulsion under strong confinements? Can they propel faster within confined environments 
in comparison to that within unconfined surroundings? 

To address some of these questions, we investigate squirmer behavior in cylindrical microchannels. In particular, we develop 
a theoretical model of swimmer motion within a cylindrical capillary, which predicts its propulsion as a function of confinement and 
its swimming strength. Furthermore, we perform simulations of microswimmer locomotion in a tube using a squirmer model
\cite{Lighthill_SMB_1952,Blake_SEA_1971,Alarcon_MCP_2017}. In both the theoretical model and simulations, periodic 
boundary conditions along the capillary axis as well as an impenetrable wall at the ends of the tube, representing a dead end, 
are considered. Simulation results confirm that the analytical model properly captures the qualitative behavior of 
a squirmer inside a cylindrical microchannel. 
In the squirmer model with fixed surface slip velocity, the propulsion force increases with increasing confinement, while the swimming velocity 
always remains smaller than the swimmer speed for unconfined conditions, suggesting that the drag on the swimmer increases 
faster than the propulsion force with confinement. The model shows that a swimmer is expected to be able to move even 
under very strong confinements, though an increasing power with increasing confinement is required. The analytical model 
does not consider details of the local flow field generated by the swimmer, while in simulations, the squirmer model 
allows the imposition of different local flow fields, corresponding to pusher, neutral, and puller swimmers \cite{Alarcon_MCP_2017}. 
The swimming velocities in long channels only differ slightly among these three swimming modes. However, the propulsion forces 
near dead ends of the tube differ substantially. These results help us better understand swimming behavior under confinement, 
including tether pulling from fluid vesicles by encapsulated microswimmers \cite{Takatori_ACF_2020,Vutukuri_APV_2020,Le_Nagard_EBV_2022,Iyer_SAV_2022}.

The article is organized as follows. Section \ref{sec:model_description} provides all necessary details about the employed methods 
and models. Section \ref{sec:analytic_description_of_the_micro_swimmer_in_a_confined_environment} presents the analytical model of 
swimmer propulsion in a capillary. The corresponding simulations using the squirmer model are presented in Section \ref{sec:simulating_micro_swimmer_in_a_confined_environment}, 
and compared with the analytical model. We discuss the results and shortly conclude in Section \ref{discussion}.

\section{Methods and models}
\label{sec:model_description}

\subsection{Modeling fluid flow}
\label{subsec:fluid_model}

To model fluid flow, we employ the dissipative particle dynamics (DPD) method \cite{Hoogerbrugge_SMH_1992,Espanol_SMO_1995,Espanol_DPD_2017}, a mesoscopic
simulation technique. In DPD, the fluid is represented by a collection of particles which correspond to small fluid volumes and interact through three types of
pairwise forces. Thus, the force on fluid particle $i$ is given by 
\begin{equation}
    {\bf F}_{i} = \sum_{j \neq i} {\bf F}^C_{ij} + {\bf F}^D_{ij} +  {\bf F}^R_{ij},
\end{equation}
where the sum runs over particles $j$ that are neighbors of the particle $i$, and includes conservative (${\bf F}^C$), dissipative (${\bf F}^D$), and
random (${\bf F}^R$) forces. The conservative force ${\bf F}^C_{ij} = a_{ij} \omega^C(r_{ij}) \hat{\bf r}_{ij}$ supplies a soft repulsion between DPD particles
with a strength coefficient $a_{ij}$, where ${\bf r}_{ij} = {\bf r}_i - {\bf r}_j$, $r_{ij} = |{\bf r}_{ij}|$, $\hat{\bf r}_{ij} = {\bf r}_{ij}/r_{ij}$. 
We introduce a weight function $\omega$ of the general form 
\begin{equation}
    \omega(r) = \left\{
        \begin{array}{r@{,\quad\text{for}\quad}l}
            (1 - r/r_c)^s & r \leq r_c,  \\
            0 & r > r_c,
        \end{array} \right. 
\end{equation}
where $r_c$ is the cutoff radius beyond which all interactions vanish, and $s$ is an exponent to modify the weight function.
The other two forces are ${\bf F}^D_{ij} = -\gamma \omega^D(r_{ij}) (\hat{\bf r}_{ij} \cdot {\bf v}_{ij}) \hat{\bf r}_{ij}$ and
${\bf F}^R_{ij} = \sigma \omega^R(r_{ij}) \xi_{ij} \hat{\bf r}_{ij} / \sqrt{\Delta t}$, where $\gamma$ and $\sigma$ are the force strength coefficients,
${\bf v}_{ij} = {\bf v}_i - {\bf v}_j$, $\xi_{ij} = \xi_{ji}$ is a Gaussian random variable with zero mean and unit variance, and $\Delta t$ is the time
step. The fluctuation-dissipation theorem relates these weight functions and the force coefficients as \cite{Espanol_SMO_1995} 
\begin{equation}
\label{eq:gamma}
    \omega^D(r) = [\omega^R(r)]^2 = \omega(r), \quad \quad \sigma^2 = 2 \gamma k_B T.
\end{equation}
Thus, the pair of dissipative and random forces corresponds to a thermostat which maintains a constant temperature $T$ in the simulated fluid.

\subsection{Swimmer model: Squirmer}
\label{subsection:swimmer_model}

A microswimmer is modeled by a triangulated network of particles which are placed homogeneously on the swimmer surface. In order to maintain a nearly
rigid shape of the squirmer, the triangulated network of particles incorporates shear elasticity, bending rigidity, and constraints for the surface
area and enclosed volume \cite{Fedosov_RBC_2010,Fedosov_MBF_2014}. Bonds between neighboring vertices of the network supply shear elasticity and
are described by the potential $U_\text{bond} = U_\text{WLC} + U_\text{POW}$ \cite{Fedosov_RBC_2010} with
\begin{equation}
    U_\text{WLC} = \frac{k_BT l_m}{4p}\frac{3x^2-2x^3}{1-x}, \quad \quad U_\text{POW} = \frac{k_p}{l},	
\end{equation}
where $x = l/l_m \in (0,1)$, $l$ is the spring length, $l_m$ the maximum spring extension, $p$ the persistence length, and $k_p$ the repulsive
force coefficient. The bending resistance is implemented through the discretization of the Helfrich bending energy
\cite{Helfrich_EPB_1973,gompper2004triangulated} as
\begin{equation}
    U_\text{bend}= \frac{\kappa}{2} \sum_i \sigma_{i} (H_i-H_0^i)^2,
\label{helfirch_eq_des}
\end{equation}
where $\kappa$ is the bending rigidity, $H_i = {\bf n}_{i} \cdot \sum_{j(i)} \sigma_{i j} {\bf r}_{i j}/(\sigma_{i}r_{i j})$ is the discretized mean curvature at
vertex $i$,  ${\bf n}_{i}$ is the unit normal at the membrane vertex $i$, $\sigma_{i} =\sum_{j(i)} \sigma_{i j}r_{i j}/4$ is the area corresponding
to vertex $i$ (the area of the dual cell), $j(i)$ corresponds to all vertices linked to vertex $i$, and $\sigma_{i j}=r_{ij}(\cot\theta_1+ \cot\theta_2)/2$
is the length of the bond in the dual lattice, where $\theta_1$ and $\theta_2$ are the angles at the two vertices opposite to the edge $ij$ in the dihedral.
Finally, $H_0^i$ is the spontaneous curvature at vertex $i$, which can be used to implement shapes with non-constant local curvatures (e.g. spheroidal surfaces).

The area and volume conservation constraints are introduced through the potentials \cite{Fedosov_RBC_2010}
\begin{equation}\label{eq:av}
  U_\text{area} = \frac{k_a(A-A_0^{tot})^2}{2A_0^{tot}} + \sum_{m\in 1...N_t}{\frac{k_d(A_m-A^m_0)^2}{2A^m_0}},
\end{equation}
\begin{equation}  
U_\text{vol} =  \frac{k_v(V-V_0^{tot})^2}{2V_0^{tot}},
\end{equation}
where $A$ is the instantaneous area of the membrane, $A_0^{tot}$ the targeted global area, $A_m$ the area of the m-th triangle, $A^m_0$ the targeted
area of the m-th triangle determined by the triangular mesh on the squirmer surface, $V$ the instantaneous membrane volume, and $V_0^{tot}$ the targeted volume. Furthermore, $k_a$, $k_d$, and $k_v$ are the global
area, local area and volume constraint coefficients, respectively. $N_t$ is the number of triangles on the triangulated surface.

The swimmer is embedded in a DPD fluid. DPD particles are also placed inside the swimmer. The swimmer surface,
described by a membrane, is impenetrable for fluid particles and separates particles inside and outside of the swimmer. Note that the dissipative and random forces 
between inside and outside fluid particles are turned off, while the conservative force is used to keep homogeneous fluid pressure across the membrane. 
To prevent fluid particles from crossing the membrane, their reflection from both sides of the membrane is implemented following a rule
\cite{Revenga_BMI_1998}, in which particle positions are updated according to specular reflection, while particle velocities follow the bounce-back reflection.

Propulsion of the squirmer is implemented through a slip velocity at its surface according to the squirmer model \cite{Alarcon_MCP_2017} as
\begin{equation}
\label{eq:surface_velocity}
    {\bf u}_\text{surf} = \left( B_1 \sin(\theta) + B_2 \sin(\theta)\cos(\theta)\right) {\bf t_{\theta}},
\end{equation}
where ${\bf t_{\theta}}$ is the local tangential vector in $\theta$ direction at the squirmer surface and $\theta$ is the angle between the orientation of the swimmer and the vector from 
its center of mass to a local position at the surface. The coefficient $B_1$ defines propulsion strength of the swimmer and $B_2$ introduces fore-and-aft asymmetry 
into the slip-velocity field. A squirmer swims in bulk fluid with a velocity $v_0=\frac{2}{3} B_1$. We use the ratio $\beta = B_2/B_1$ to capture the modality. For $\beta >0$ the squirmer is called a puller, for $\beta = 0$ it is neutral and for $\beta <0$ it is a pusher. Adaptation of Eq.~(\ref{eq:surface_velocity}) to a spheroidal shape is described in Appendix A. To enforce the slip 
velocity at the swimmer surface, dissipative interaction between swimmer particles and those of outer fluid is modified as   
\begin{equation}
    {\bf F}_{ij}^D = -\gamma \omega^D(r_{ij})(\hat{\bf r}_{ij}\cdot{\bf v}_{ij}^\ast)\hat{\bf r}_{ij}, \quad  \quad 
    {\bf v}_{ij}^\ast = {\bf v}_i - {\bf v}_j + {\bf u}_\text{surf}^i,
\end{equation}
where ${\bf u}_\text{surf}^i$ is the slip velocity at the position of swimmer particle $i$, while $j$ represents an outer-fluid particle. 

\subsection{Model of solid walls}
\label{subsection:confinement_model}

Solid walls are represented by frozen DPD particles which have the same number density and interactions as fluid particles. The frozen wall particles facilitate
no-slip boundary conditions (BCs) at the walls through dissipative interactions with fluid particles. To prevent fluid particles from penetrating the walls, 
reflective surfaces are added at the fluid-solid interface, where bounce-back reflection of fluid particles is performed \cite{Revenga_BMI_1998}. 
Note that the bounce-back reflection also contributes to the enforcement of no-slip BCs at the walls. 

\subsection{Simulation parameters}
\label{subsec:simuation_implementation}

In most simulations, the swimmer has a spherical shape with a radius $r_{sq}=5r_c$, and is represented by the squirmer 
model with $N=3000$ discretization particles on its surface. To maintain a non-deformable spherical geometry, the targeted membrane area 
$A_0^{tot}$ is set $5\%$ smaller than $4\pi r_{sq}^2$ and the targeted volume $V_0^{tot}$ $5\%$ larger than $4\pi r_{sq}^3/3$, such that the 
surface is under considerable tension. The confinement is modeled by a capillary with a radius $R_{cap}$ using either periodic BCs in the $z$ direction
or solid-wall BCs at both ends of the tube. To constrain the swimmer position at the capillary center and its swimming orientation along the $z$ axis,
two springs are attached to the front and rear portions of the swimmer, represented by about $300$ surface particles with maximum and minimum 
$z$ coordinates, respectively. Equilibrium position of these springs is set to (0, 0, *), such that the springs control the swimmer position and 
orientation only in the $x$ and $y$ directions, while the motion along the $z$ direction remains unaffected. In addition to spherically-shaped 
squirmers, a number of simulations are also performed for a prolate spheroid ($N=1038$ particles) with semi-minor axes $b_x = b_y = 3r_c$ and 
a semi-major axis $b_z=6r_c$.

The fluid is modelled by a collection of DPD particles with a number density $n=5/r_c^3$ ($r_c=1.0$ is selected in simulations). The DPD force 
coefficients for interactions between fluid particles are given by $a=80 k_B T / r_c$, $\gamma=50 \sqrt{m k_B T}/ r_c$, $s=1$ for $\omega^C$ and $s=0.1$ for $\omega^D$ and $\omega^R$, where $m=1$
is the particle mass and $k_B T = 1$ is the unit of energy. This results in a fluid dynamic viscosity of $\eta=77.5 \sqrt{m k_B T}/ r_c^2$. Coupling 
between the spherical squirmer and fluid particles assumes $a^{sf}=0$ and $\gamma^{sf}=59.7 \sqrt{m k_B T}/ r_c$ with $r_c^{sf}=0.8 r_c$. The value of $\gamma^{sf}$
is computed such that the imposed slip BCs at the squirmer surface are obtained \cite{Fedosov_RBC_2010}. For the spheroidal swimmer, $\gamma^{sf}= 106.1 \sqrt{m k_B T}/ r_c$
because the surface density of squirmer particles is different from that for the spherical swimmer. For simulations with a finer fluid resolution, 
the density is set to $n=20/r_c^3$, while the cutoff radius is $r'_c= 0.6 r_c$ to reduce the overall computational cost. In this case, the dynamic viscosity 
becomes $\eta=84.7 \sqrt{m k_B T}/ r_c^2$, which is close to that in simulations with $n=5/r_c^3$.
The time step is $\Delta t = 0.0025 r_c \sqrt{m / k_B T}$, and the center of mass of the squirmer is sampled every 100 time steps. The velocity is computed using the distance that the swimmer covers within 800 time steps. Error bars represent the standard deviation of these samples.

\begin{figure}[t]
\centering
  \includegraphics[width=\linewidth]{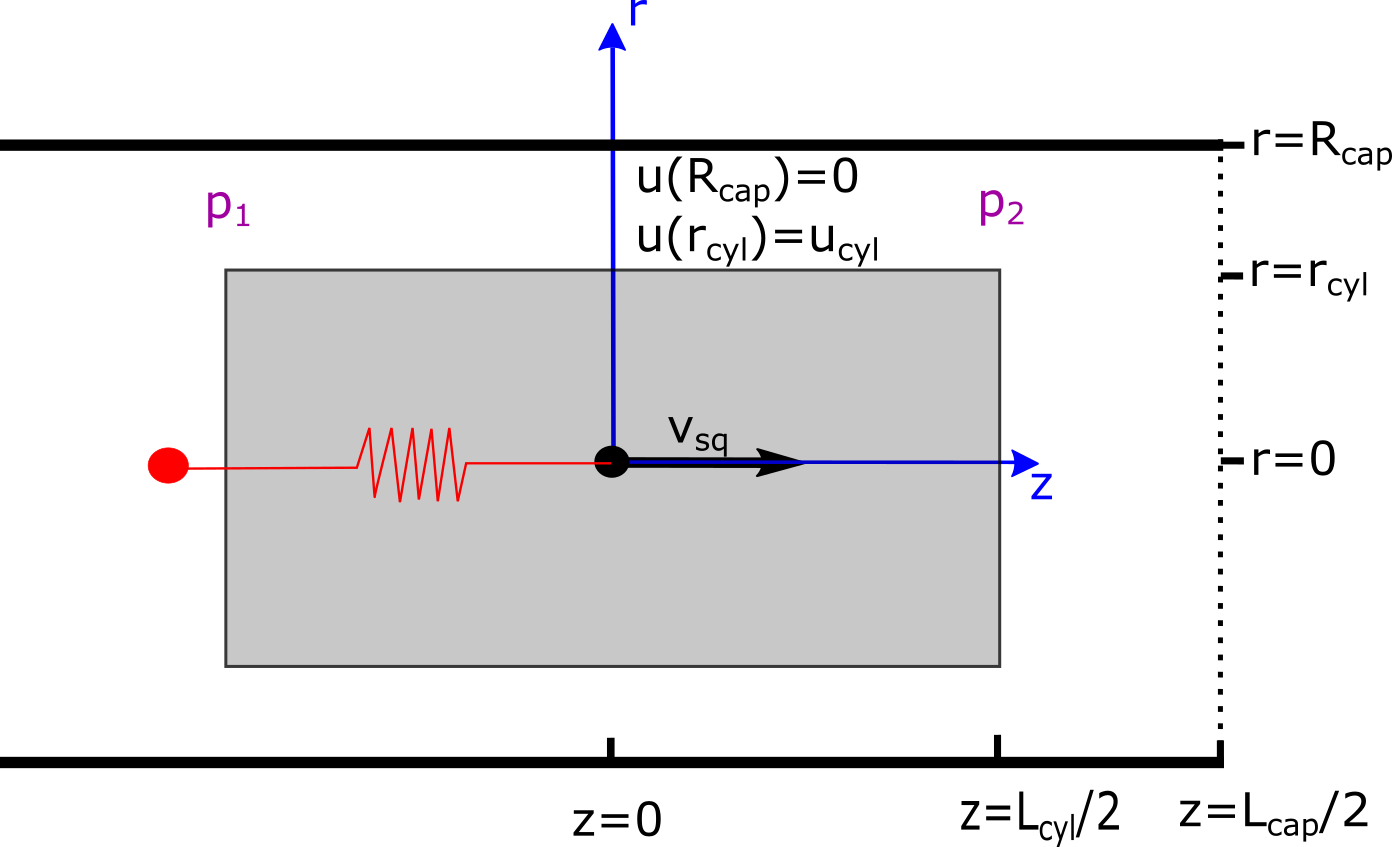}
  \caption{Schematic of a cylindrical swimmer in a periodic cylindrical capillary. The swimmer has a radius $r_{cyl}$ and a length $L_{cyl}$, while the 
  capillary is characterized by a radius $R_{cap}$ and a length $L_{cap}$. $p_1$ and $p_2$ denote pressures at the back and the front of the swimmer, respectively. 
  A free swimmer moves with a velocity $v_{sq}$. The swimmer's center of mass (the black point at $z=0$) can also be held by a spring (red line), in order to
  measure its propulsion force. In that case, the boundary conditions of the flow profile are given by the slip velocity $u_{cyl}$ at the swimmer surface and the no-slip condition at the capillary wall.}
  \label{fig:ana_sketch}
\end{figure}

\section{Analytical model of squirmer propulsion in cylindrical microchannels}
\label{sec:analytic_description_of_the_micro_swimmer_in_a_confined_environment}

For an analytical description of squirmer propulsion under confinement, we consider a cylindrical swimmer shape with a radius $r_{cyl}$ and a length $L_{cyl}$,
see a sketch in Fig.~\ref{fig:ana_sketch}. Swimmer propulsion is facilitated by a constant slip velocity $u_{cyl}$ at the cylinder jacket. 
The confinement is represented by a capillary with a radius $R_{cap}$ and a length $L_{cap}$, which can either be closed at its both ends or have periodic BCs 
along the tube axis (see Fig.~\ref{fig:ana_sketch}). The surface of the capillary is subject to no-slip BCs for fluid flow. 
$p_1$ and $p_2$ correspond to fluid pressures at the back and front ends of the swimmer, respectively. The swimmer axis is assumed to always be parallel to the tube axis and at the centerline of the capillary. We are interested in the velocity 
$v_{sq}$ of the swimmer, when its propulsion force is balanced by the drag due to surrounding fluid flow. Furthermore, to measure the propulsion force 
of the swimmer, its center of mass can be pinned by a harmonic spring (see Fig.~\ref{fig:ana_sketch}), such that the retaining force of the spring equals 
the propulsion force of the swimmer. 
To make this model analytically tractable, we neglect the effect of local fluid flow near its back and front ends on the propulsion, i.e., we assume that fluid radial velocity is zero everywhere, and consider only fluid flow through a cross-section of the tube.
This approximation is valid for long swimmers ($L_{cyl} \gg r_{cyl}$).
Under the assumption that the microswimmer moves at low Reynolds numbers, fluid flow in this system can be described by 
the incompressible Stokes equation \cite{Stone_PMD_1996} $\eta\bigtriangledown^2{\bf u} = \bigtriangledown p$, where $\eta$ is the dynamic viscosity. 
In cylindrical coordinates, the general solution of this equation for fluid flow through a cross-section is given by
\begin{equation}
    u_z(r)=\frac{\bigtriangledown p}{4\eta}r^2 + c_1 \ln(r) + c_2,
    \label{eq:u_z_analytical_general}
\end{equation}
where $c_1$ and $c_2$ are constants that need to be determined. Explicit analytical expressions for the parameters (e.g. $c_1$ and $c_2$) of the 
analytical model are given in Appendix B for different conditions.  

\begin{figure*}[!th]
  \centering
  \subfloat{\includegraphics[width=0.8\linewidth]{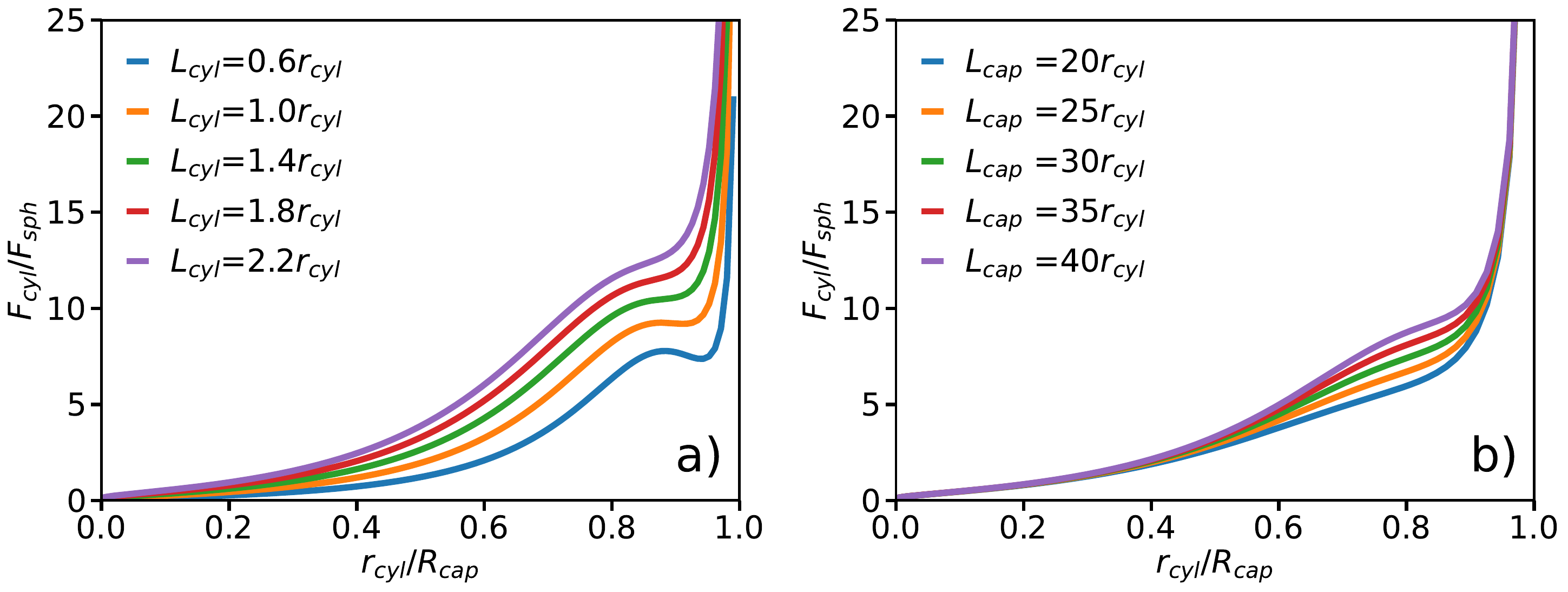}} \\
  \subfloat{\includegraphics[width=0.8\linewidth]{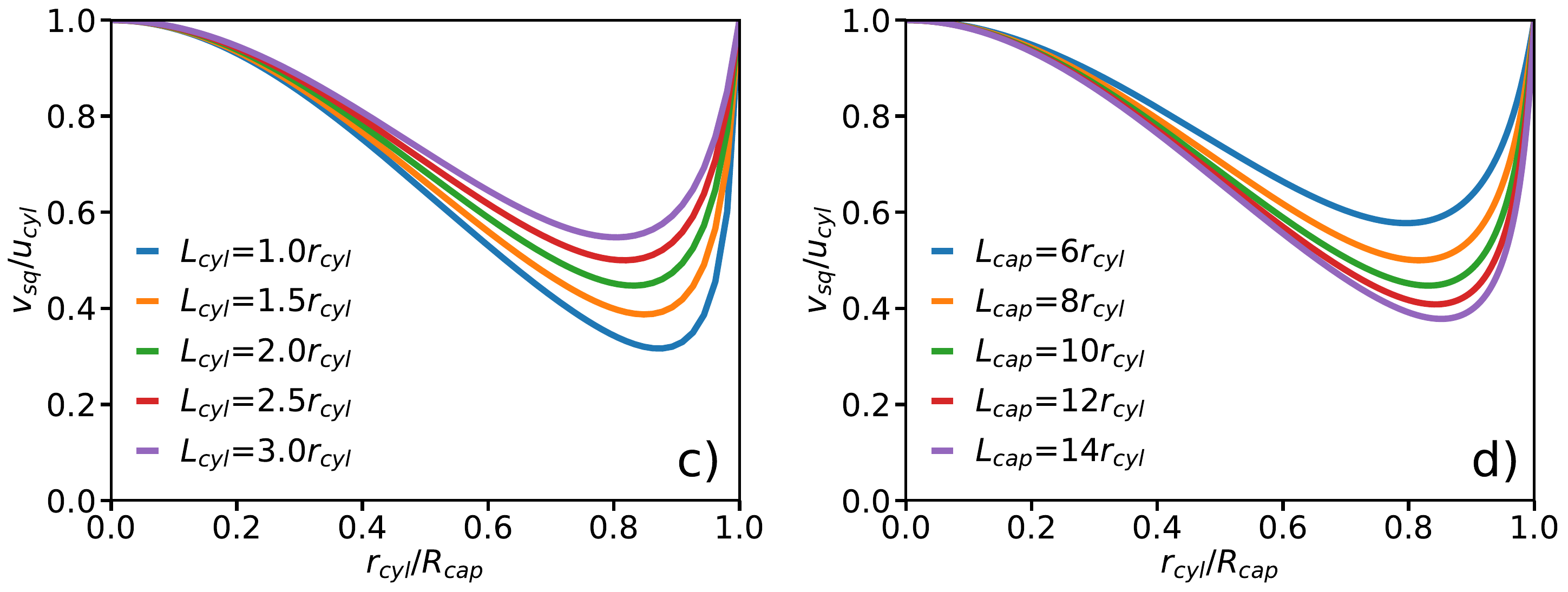}} \\
  \caption{Analytical model for a cylindrical swimmer in a periodic capillary. (a-b) Propulsion force of a spring-pinned swimmer as a function 
  of confinement for (a) various swimmer lengths with $L_{cap}=60r_{cyl}$ and (b) different capillary lengths with $L_{cyl}= 2r_{cyl}$. The confinement 
  $D = r_{cyl} / R_{cap}$ is varied by changing $R_{cap}$ and the force is normalized by $F_{sph} = 6\pi\eta r_{cyl} u_{cyl}$. (c-d) Velocity of a freely moving 
  cylindrical swimmer as a function of confinement for (c) different swimmer lengths with $L_{cap} = 10r_{cyl}$ and (d) various capillary lengths with 
  $L_{cyl} = 2r_{cyl}$.}
  \label{fig:oP_ForceVelVsConf}
\end{figure*}

\subsection{Squirmer in a periodic capillary}
\label{subsec:swimming in the open pipe}

When the swimmer's center of mass is pinned by a spring, $v_{sq}=0$ at steady state. The cross-sectional 
flow profiles can then be derived (i) through the gap between the swimmer and the capillary, and (ii) through the tube away from the swimmer.
The pressure gradient along the gap is 
given by $\bigtriangledown p_{gap}=(p_2-p_1)/L_{cyl}=\Delta p/L_{cyl}$, while the pressure gradient along the swimmer-free section of the capillary 
is $\bigtriangledown p_{cap}=(p_1-p_2)/(L_{cap}-L_{cyl})=-\Delta p/(L_{cap}-L_{cyl})$ due to periodic BCs along the tube axis. Then, the corresponding flow profiles are
\begin{align}
    u_{gap}(r) = \ &\frac{\Delta p R_{cap}^2}{4\eta L_{cyl}}\left(\frac{r^2}{R_{cap}^2} - 1\right) + \cr
     &\frac{u_{cyl} + 
    (\Delta p R_{cap}^2)/(4\eta L_{cyl}) \cdot (1-D^2)}{\ln (D)}\ln \left(\frac{r}{R_{cap}} \right), \label{eq:ugap} \\
    u_{cap}(r) = \ &\frac{\Delta p}{4\eta (L_{cap}-L_{cyl})}(R_{cap}^2-r^2),
\end{align}
where $D = r_{cyl} / R_{cap}$ is the confinement. Here, we used BCs at the capillary surface with $u(R_{cap})=0$ and at the lateral surface of the swimmer with $u_{gap}(r_{cyl})= u_{cyl}$. 
To determine the pressure drop $\Delta p$, we use mass conservation which requires the volumetric flow rate through the gap to be equal to the flow rate through 
the swimmer-free section of the tube. An integral form of mass conservation reads $\int_0^{2\pi}d\phi\int_{r_{cyl}}^{R_{cap}}dr \; r \,u_{gap}(r) = 
\int_0^{2\pi}d\phi\int_0^{R_{cap}} dr \;r \, u_{cap}(r)$, allowing us to compute $\Delta p$ and fully determine the flow profiles above.

To calculate the propulsion force ${\bf F}_{cyl}$, the fluid stress tensor ${\bm \sigma}$ has to be integrated over the microswimmer surface $S$, i.e.  
${\bf F}_{cyl} = \int_S {\bf n}\cdot \boldsymbol{\sigma} dS$ with surface normal ${\bf n}$. In cylindrical coordinates with the assumption 
that $u_r = u_\phi = 0$, the stress tensor becomes 
\begin{equation}
    \boldsymbol{\sigma} = \begin{pmatrix}
                -p & 0 & \eta \frac{\partial u_z}{\partial r}\cos(\phi) \\
                0 & -p & \eta \frac{\partial u_z}{\partial r}\sin(\phi)\\
                \eta \frac{\partial u_z}{\partial r}\cos(\phi) & \eta \frac{\partial u_z}{\partial r}\sin(\phi) & -p
    \end{pmatrix},
\end{equation}
where $p$ is the hydrostatic pressure. Integration of fluid stresses over the cylindrical surface leads to
\begin{equation}
\label{eq:F_sigma}
    {\bf F}_{cyl} = \left(2\pi r_{cyl}  L_{cyl} \eta \frac{\partial u_z}{\partial r}\Big|_{r=r_{cyl}} - \pi r_{cyl}^2 \Delta p \right) {\bf e}_z =
    2\pi  L_{cyl} \eta c_1 {\bf e}_z. 
\end{equation}
Insertion of the solution $u_{gap}(r)$ from Eq.~(\ref{eq:ugap}) results in the propulsion force
\begin{equation}
    {\bf F}_{cyl} = 2 \pi L_{cyl} \eta \frac{u_{cyl} + (\Delta p R_{cap}^2)/(4\eta L_{cyl})\cdot(1-D^2)}{\ln(D)} {\bf e}_z.
\label{eq:F_sigma_open_pipe}    
\end{equation}
As the pressure drop $\Delta p$ is proportional to $1/ R_{cap}^2$, the propulsion force in Eq.~(\ref{eq:F_sigma_open_pipe}) becomes 
a function of the confinement $D=r_{cyl} / R_{cap}$. The force ${\bf F}_{cyl}$ is displayed in Fig.~\ref{fig:oP_ForceVelVsConf}(a-b) for different swimmer and capillary lengths. 
It increases with increasing confinement, and diverges as $D \to 1$. Note that here the analytical model does not assume any limit 
on the generated propulsion force (see a discussion in Section~\ref{subsec:microswimmer_with_fixed_propulsion_force}), while forces generated by realistic microswimmers would clearly have a finite upper bound. In the limit of $D \to 0$, the propulsion force vanishes,
suggesting that the swimmer can move without any force generation in an unconfined situation. This is an artifact of model 
assumptions, where the shear rate $\frac{\partial u_z}{\partial r}$ at the cylinder jacket vanishes as $R_{cap} \to \infty$ (or $D \to 0$), 
so that the shear stress on the cylinder also disappears.   

For $L_{cyl}/L_{cap} \ll 1$, a local maximum in ${\bf F}_{cyl}$ emerges at strong confinements [see Fig.~\ref{fig:oP_ForceVelVsConf}(a)].
The force expression in Eq.~(\ref{eq:F_sigma_open_pipe}) has one component that is dependent on the pressure difference and another that is not. While the latter is responsible for the divergence at $D \to 1$, the former has a local maximum corresponding to the maximum in the pressure difference. The  pressure gradient is increasing for smaller $L_{cyl}$ which explains the emergence of the local maximum only for small values of $L_{cyl}$.
Figure \ref{fig:oP_ForceVelVsConf}(b) 
shows that the propulsion force is also affected by the capillary length $L_{cap}$ for an intermediate range of confinements, while the effect of $L_{cap}$ can be neglected 
for low and high confinements. 

For a free-swimming squirmer (i.e. no pinning spring) moving along the channel center line,  
the BCs at the swimmer surface are modified 
as $u_{gap}(r_{cyl}) =u_{cyl}+v_{sq}$. Here, an additional condition is that the swimmer is force free, i.e. ${\bf F}_{cyl} = 0$,
such that the propulsion force balances the drag force on the swimmer. With the force given in Eq.~(\ref{eq:F_sigma}),
we can conclude that $c_1=0$ in the solution for the velocity profile within the gap. Furthermore, the equation for
mass conservation changes to $\int_{0}^{2\pi} d\phi \int_{r_{cyl}}^{R_{cap}} dr \, r \, u_{gap}(r) + \pi r_{cyl}^2v_{sq} 
= \int_0^{2\pi} d\phi \int_0^{R_{cap}} dr \, r \, u_{cap}(r)$. As a result, we obtain 
the swimming velocity
\begin{equation}
    v_{sq}=-u_{cyl} - \frac{\Delta p R_{cap}^2}{4\eta L_{cyl}}(1-D^2).
\label{eq:v_anal_open_pipe}    
\end{equation}

Figure \ref{fig:oP_ForceVelVsConf}(c-d) shows the swimmer velocity as a function of confinement $D$ for
different swimmer and capillary lengths. In the limit of $r_{cyl}/R_{cap}\rightarrow 0$ or
$r_{cyl}/R_{cap}\rightarrow 1$, the swimmer velocity is equal to the slip velocity. 
With increasing confinement, the velocity first decreases, reaches a minimum, and then 
increases again. Longer swimmers propel with larger speeds [see Fig.~\ref{fig:oP_ForceVelVsConf}(c)],
as they generate larger propulsion forces. An increase in $L_{cap}$ leads to a reduction 
in the swimming speed, as shown in Fig.~\ref{fig:oP_ForceVelVsConf}(d). These predictions are consistent 
with previous simulation results for confined spherical squirmers \cite{Zhu_LRN_2013}, where a decrease in 
swimming velocity was found for increasing confinements within the range from $0.2$ to $0.5$. This behavior can be understood as follows. In the limit $D\to 1$, there cannot be any shear gradient in the gap, so that $v_{sq} = -u_{cyl}$. For smaller D, the propulsion force is finite, but is has to work against a friction force of the fluid column in the tube, which increases linearly with the tube length. Thus, $v_{sq}\sim 1/L_{cap}$ at the minimum of $v_{sq}$.

\subsection{Squirmer with a fixed propulsion force}
\label{subsec:microswimmer_with_fixed_propulsion_force}

\begin{figure}[!th]
    \centering
    \includegraphics[width=0.8\linewidth]{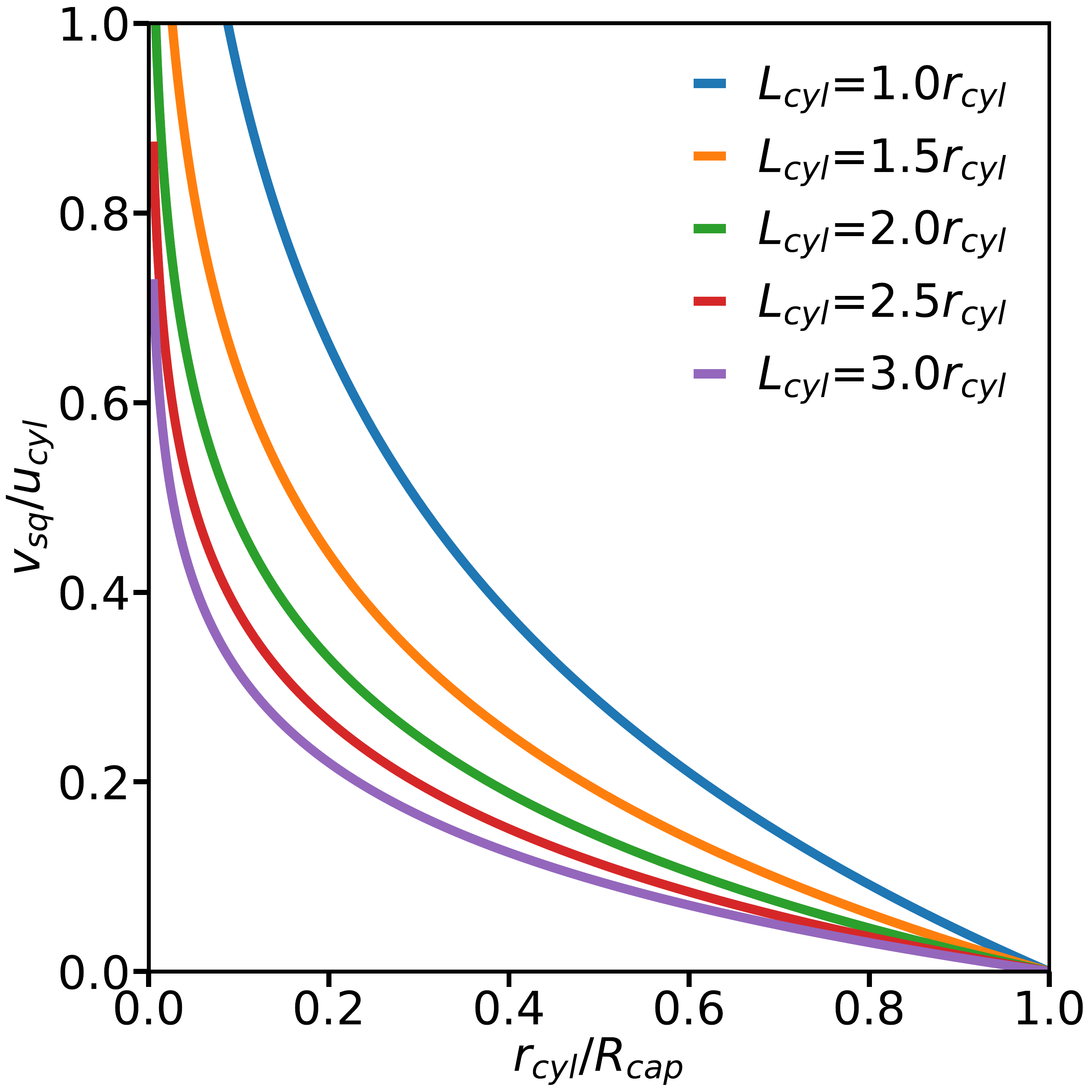}
    \caption{Velocity of a freely moving cylindrical swimmer with a fixed propulsion force of $F_{max}=0.14 F_{sph}$ ($F_{sph} = 6\pi\eta r_{cyl} u_{cyl}$) as a function of confinement for different swimmer lengths. The confinement $r_{cyl}/R_{cap}$ is varied by changing $R_{cap}$.}
    \label{fig:frSoP_fixedForce}
\end{figure}

The results in Section~\ref{subsec:swimming in the open pipe} assume a fixed surface slip velocity, from which the required propulsion force is then derived.
We now assume a fixed propulsion force $F_{cyl} = F_{max}$ for the spring-pinned case instead. The combination of Eq.~(\ref{eq:F_sigma}) with the BCs for $u_{gap}$ results in a confinement-dependent surface velocity
\begin{equation}
    \label{eq:surVel_constForce}
    u_{cyl} = \frac{1}{4\eta L_{cyl}}\cdot \big(\frac{2F_{max}\ln (D)}{\pi} - \Delta p R_{cap}^2(1-D^2) \big).
\end{equation}
When this surface velocity is used for the free-swimming case by inserting it into Eq.~(\ref{eq:v_anal_open_pipe}), the swimming velocity becomes
\begin{equation}
    \label{eq:v_anal_fixedForec}
    v_{sq} = - \frac{F_{max} \ln (D)}{2\pi\eta\L_{cyl}}.
\end{equation}
This result is independent of $L_{cap}$. The dependence on $D$ is shown in Fig.~\ref{fig:frSoP_fixedForce}. As expected, the swimming velocity goes to zero for $D\to 1$ for a swimmer with a finite propulsion force. Also, the velocity diverges in the absence of confinement, which is an artifact of the theoretical model related to the fact that $F_{cyl}$ vanishes as $D\to 0$. This means that 
the model predicts zero fluid friction on a moving swimmer in the limit of $D\to 0$, suggesting that the model should not be used 
for $D \ll 1$. Furthermore, the ability of real swimmers is always limited by a finite value of the imposed surface velocity.

\begin{figure*}[!th]
    \centering
    \includegraphics[width=0.8\linewidth]{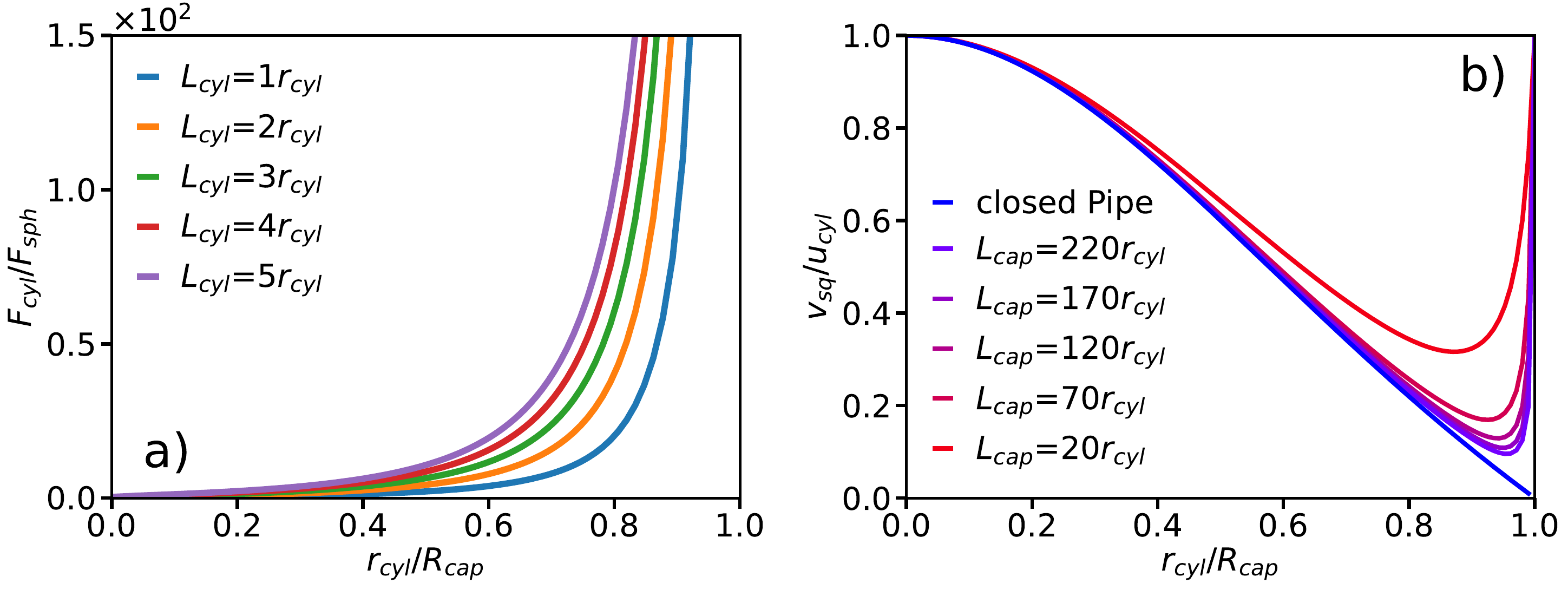}
    \caption{Analytical model for a cylindrical swimmer in a capillary. (a) Propulsion force of a spring-pinned swimmer in a closed tube as a function 
  of confinement for different swimmer lengths. The confinement $D = r_{cyl} / R_{cap}$ is varied by changing $R_{cap}$ and the force is normalized by 
  $F_{sph} = 6\pi\eta r_{cyl} u_{cyl}$. (b) Velocity of a freely moving cylindrical swimmer in a closed tube far away from the both tube ends and in an open tube with periodic BCs as a function of confinement.}
    \label{fig:fiScP-fPAL}
\end{figure*}

\subsection{Squirmer motion in a closed capillary}
\label{swimming_in_the_closed_pipe}

Another situation is when the confinement geometry is closed, representing two dead ends. For instance, this is the case for swimmers in lipid vesicles 
located at the end of formed tethers \cite{Vutukuri_APV_2020,Takatori_ACF_2020}. When the center of mass of the swimmer is tethered by a spring, mass
conservation corresponds to zero fluid flux through the gap between the swimmer and the capillary, and reads $\int_0^{2\pi}d\phi \int_0^{R_{cap}} dr \;r \, 
u_{gap}(r) = 0$. In combination with the BCs $u(R_{cap})=0$ and $u(r_{cyl})= u_{cyl}$, we can determine the parameters $c_1$, $c_2$, and $\Delta p$ in Eq.~(\ref{eq:u_z_analytical_general}), 
leading to a solution for the flow profile in the gap and the resulting propulsion force 
\begin{equation}
\label{eq:fiScP_zDeviation}
    {\bf F}_{cyl} = 2\pi L_{cyl} \eta\frac{u_{cyl}}{1+\ln(D)(1+D^2)/(1-D^2)}{\bf e}_z.
\end{equation}
The dependence of $F_{cyl}$ on $D$ is shown in Fig.~\ref{fig:fiScP-fPAL}(a) for different $L_{cyl}$. The propulsion force increases with increasing confinement, and diverges 
as $r_{cyl} / R_{cap} \rightarrow 1$. Similar to the case of the periodic tube, longer swimmers generate more thrust.

For a freely-moving swimmer in a closed tube, the mass conservation reads $\int_0^{2\pi} d\phi \int_{r_{cyl}}^{R_{cap}} dr \, r \, 
u_{gap}(r) + \pi r_{cyl}^2v_{sq} = 0$, resulting in the swimming velocity
\begin{equation}
\label{eq:frSqcP_velocity}
    v_{sq} = -u_{cyl}\cdot \frac{1-D^2}{1+D^2}.
\end{equation}
Figure \ref{fig:fiScP-fPAL}(b) shows that the velocity decreases with increasing confinement and vanishes as $D\rightarrow 1$, despite an increasing 
propulsion force in Fig.~\ref{fig:fiScP-fPAL}(a). This indicates that the drag on the swimmer increases faster than its propulsion force for increasing confinement.
When we consider the case of $D=1$, the fluid exchange between the region anterior and posterior of the swimmer vanishes. 
For an incompressible fluid, the flow through the gap must also vanish due to volume conservation. 
This also explains the different behavior in comparison to the periodic tube, where the swimmer can "push" the fluid in front and hence can swim with a non-zero velocity. 
Note that the force and velocity of the swimmer in the closed tube are independent of the capillary length, since mass conservation is only affected by the flow in the gap between the swimmer and the capillary wall.

In order to compare the analytical model for a closed tube with that of a periodic one, Fig.~\ref{fig:fiScP-fPAL} (b) presents the velocity of a cylindrical swimmer
as a function of confinement for the both models. In the limit of $L_{cap} \to \infty$, the analytical model with periodic BCs along the capillary axis converges to that for a closed
tube. As the capillary length increases, the overall resistance for fluid flow also increases, leading to a negligible volumetric flow rate within the tube 
for large capillary lengths.

A calculation based on a fixed propulsion force is similar to that for a capillary with periodic BCs in Section~\ref{subsec:microswimmer_with_fixed_propulsion_force}. Therefore, the predictions for $F_{cyl} = F_{max}$ in a closed capillary do not differ qualitatively from those in Fig.~\ref{fig:frSoP_fixedForce}.

\begin{figure}[!bh]
    \centering
    \includegraphics[width=0.8\linewidth]{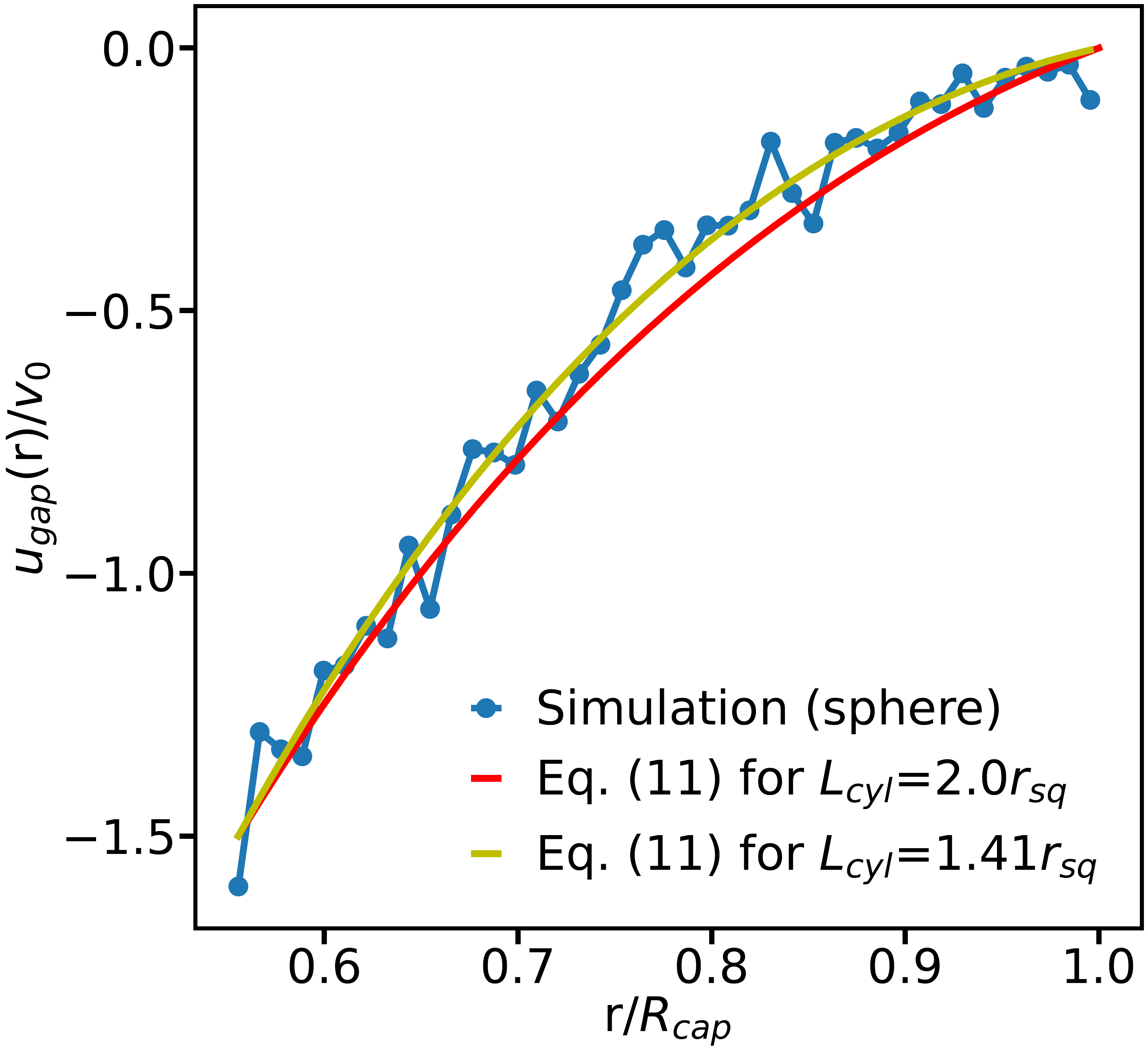}
    \caption{Flow velocity profile $u_{gap}(r)$ in the gap at the position ($z = 0$) of the center of mass of a spring-fixed spherical squirmer. The confining 
    capillary is periodic along its axis and the velocity is normalized by $v_0 = \frac{2}{3}B_1$, which is the swimming velocity of a squirmer without confinement.
    Here, the confinement is $D=r_{sq} / R_{cap} = 0.56$. The blue curve corresponds to averaged velocity profile from the simulation, while the red and 
    orange lines represent velocity profiles from the analytical solution in Eq.~(\ref{eq:ugap}) for a cylinder with a radius $r_{cyl}=r_{sq}$, $u_{cyl} = -B_1$ and 
    lengths $L_{cyl}=2r_{sq}$ and $L_{cyl}=1.41r_{sq}$, respectively.}
    \label{fig:fiSqoP_sim-radialFlowProfile}
\end{figure}

\begin{figure*}[!th]
    \centering
    \includegraphics[width=0.8\linewidth]{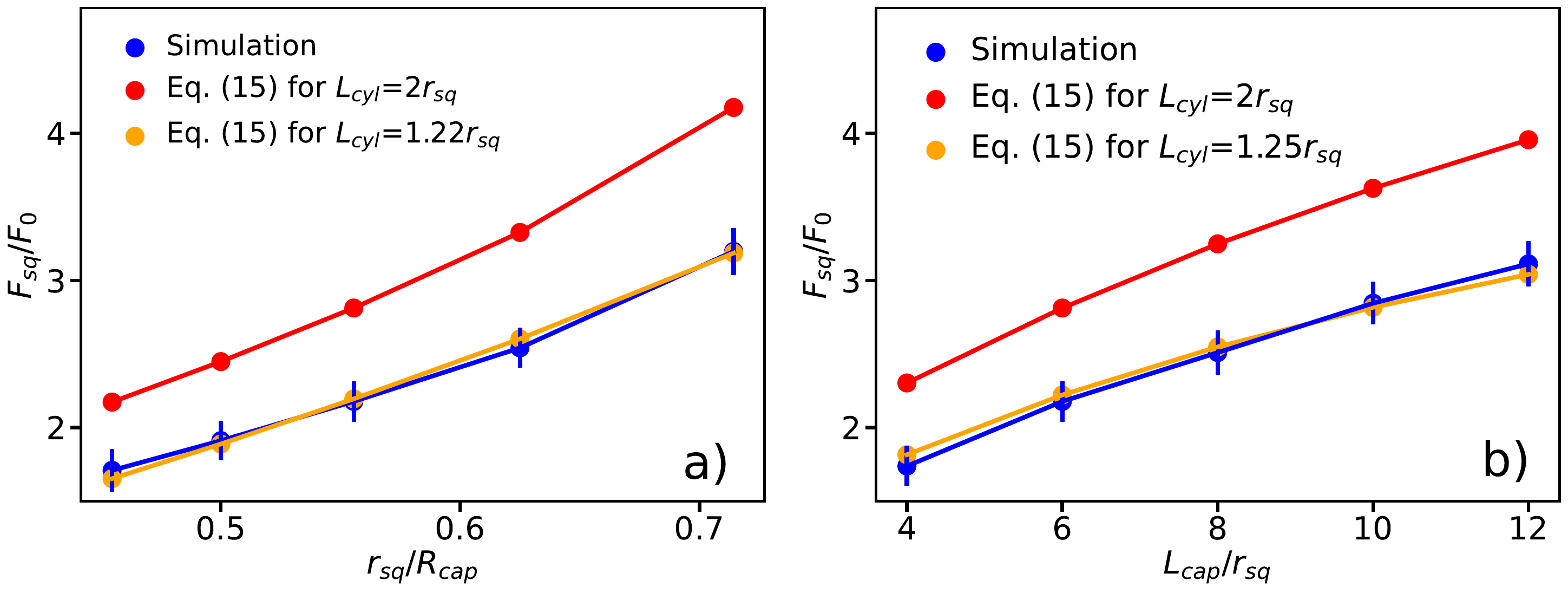}
    \includegraphics[width=0.8\linewidth]{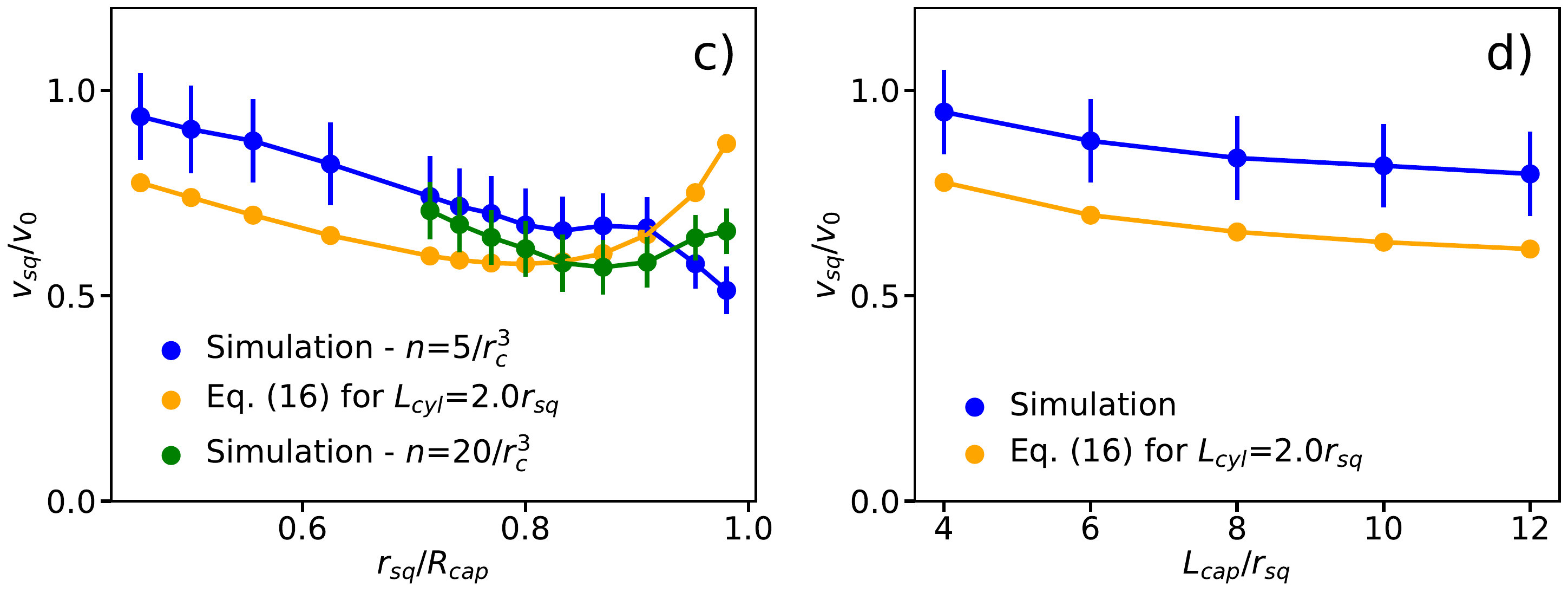}
    \caption{(a-b) Propulsion force $F_{sq}$ of a spring-fixed spherical squirmer in comparison with the analytical prediction from Eq.~(\ref{eq:F_sigma_open_pipe})
    for (a) different confinements $D=r_{sq} / R_{cap}$ with $L_{cap}=6r_{sq}$ and (b) various capillary lengths $L_{cap}$ with $D = 0.56$. 
    The red and orange lines show propulsion forces from the analytical solution in 
    Eq.~(\ref{eq:F_sigma_open_pipe}) for a cylindrical swimmer with radius $r_{cyl}=r_{sq}$, surface velocity $u_{cyl}=-B_1$ and two different lengths. The force is normalized by 
    $F_0 = 6\pi\eta r_{sq} v_0$ with $v_0 = \frac{2}{3}B_1$. (c-d) Swimming velocity of the squirmer in comparison with the analytical prediction 
    from Eq.~(\ref{eq:v_anal_open_pipe}) for (c) different confinements $D=r_{sq} / R_{cap}$ with $L_{cap}=6r_{sq}$ and (d) various capillary lengths $L_{cap}$ 
    with $D = 0.56$. The velocity is normalized by $v_0$. The orange line corresponds to the analytical solution from 
    Eq.~(\ref{eq:v_anal_open_pipe}) for a cylinder with a radius $r_{cyl}=r_{sq}$, a surface velocity $u_{cyl} = -v_0$ and a cylinder length $L_{cyl}=2r_{sq}$. Fitting the cylinder length results in the same value. The green curve in (c) 
    shows the swimming velocity at strong confinements from simulations with an increased fluid resolution (i.e. particle density  $n = 20/r_c^3$). The confinement is varied by changing $R_{cap}$. The error bars represent standard deviation 
    of simulation measurements. Periodic BCs in the $z$ direction are assumed in all cases.}
    \label{fig:fiSoP_sim-x0AR}
\end{figure*}

\begin{figure*}[!th]
    \centering
    \includegraphics[width=0.4\linewidth]{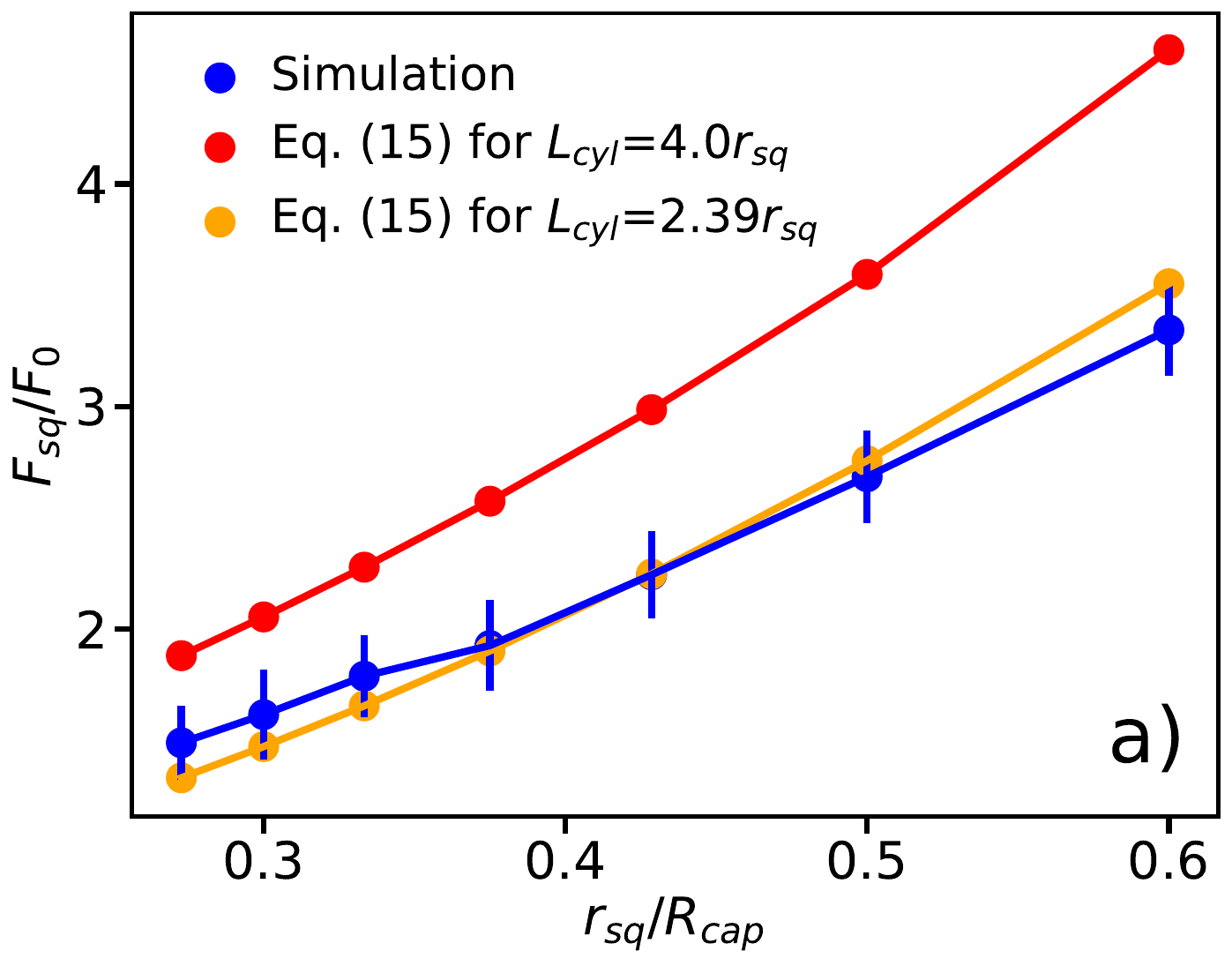}
    \includegraphics[width=0.4\linewidth]{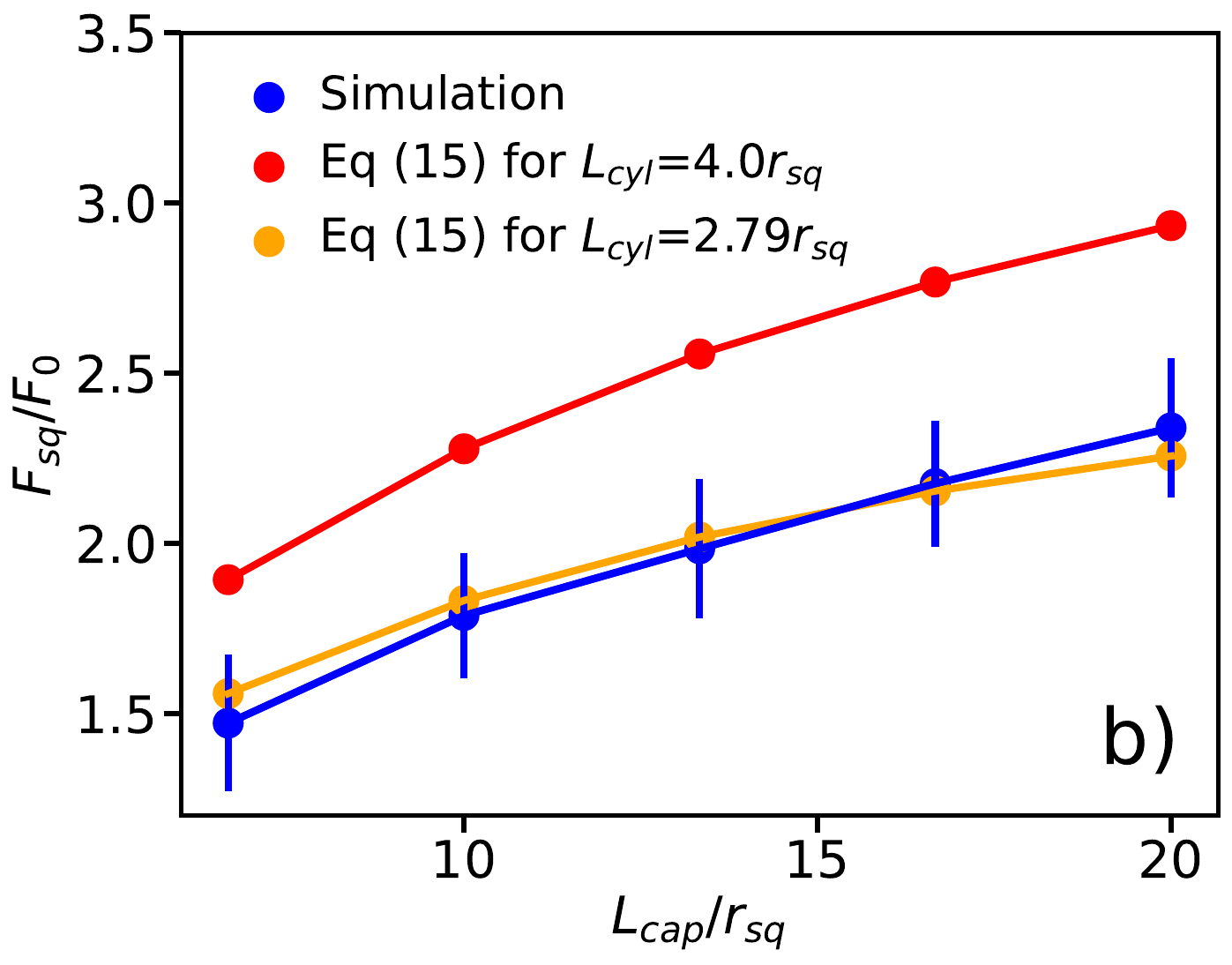}
    \caption{Propulsion force $F_{sq}$ of a spring-fixed spheroidal squirmer with $b_x=b_y$ and $b_z=2b_x$ in an open capillary in comparison with the analytical prediction from Eq.~(\ref{eq:F_sigma_open_pipe}) for (a) different confinements $ r_{sq}/r_{sq}$ with $L_{cap}=10r_{sq}$ and (b) various capillary lengths $L_{cap}$ with $D=0.33$. The red and orange lines show propulsion forces from the analytical solution in Eq.~(\ref{eq:F_sigma_open_pipe}) for a cylindrical swimmer with radius $r_{cyl}=r_{sq}$, surface velocity $u_{cyl}=-B_1$ and two different lengths $L_{cyl}$.The force is normalized by $F_0 = 6\pi\eta r_{sq} v_0$ with $v_0 = 0.83B_1$, which is the bulk velocity of a spheroidal squirmer with the given eccentricity \cite{Theers_MSM_2016}.}
    \label{fig:fiSoP_sim_spheroidal}
\end{figure*}

\section{Simulations of squirmer motion in cylindrical microchannels}
\label{sec:simulating_micro_swimmer_in_a_confined_environment}

In order to verify to which extent the analytical model captures the behavior of squirmer in microchannels, we perform numerical simulations of a swimmer in a capillary for various conditions.
In simulations, we use either spherical or spheroidal squirmer models, as described in Section~\ref{subsection:swimmer_model}.
The results in Sections~\ref{subsection:simulating_a_micro_swimmer_in_open_pipe} and \ref{subsection:simulating_a_micro_swimmer_in_an_closed_pipe} are based on a neutral squirmer model ($\beta=0$), while Section~\ref{subsection:The_affect_of_the_swimming_mode_on_the_propulsion_force}  analyses the effect of different swimming modes.

\begin{figure*}[!th]
    \includegraphics[width=0.33\linewidth]{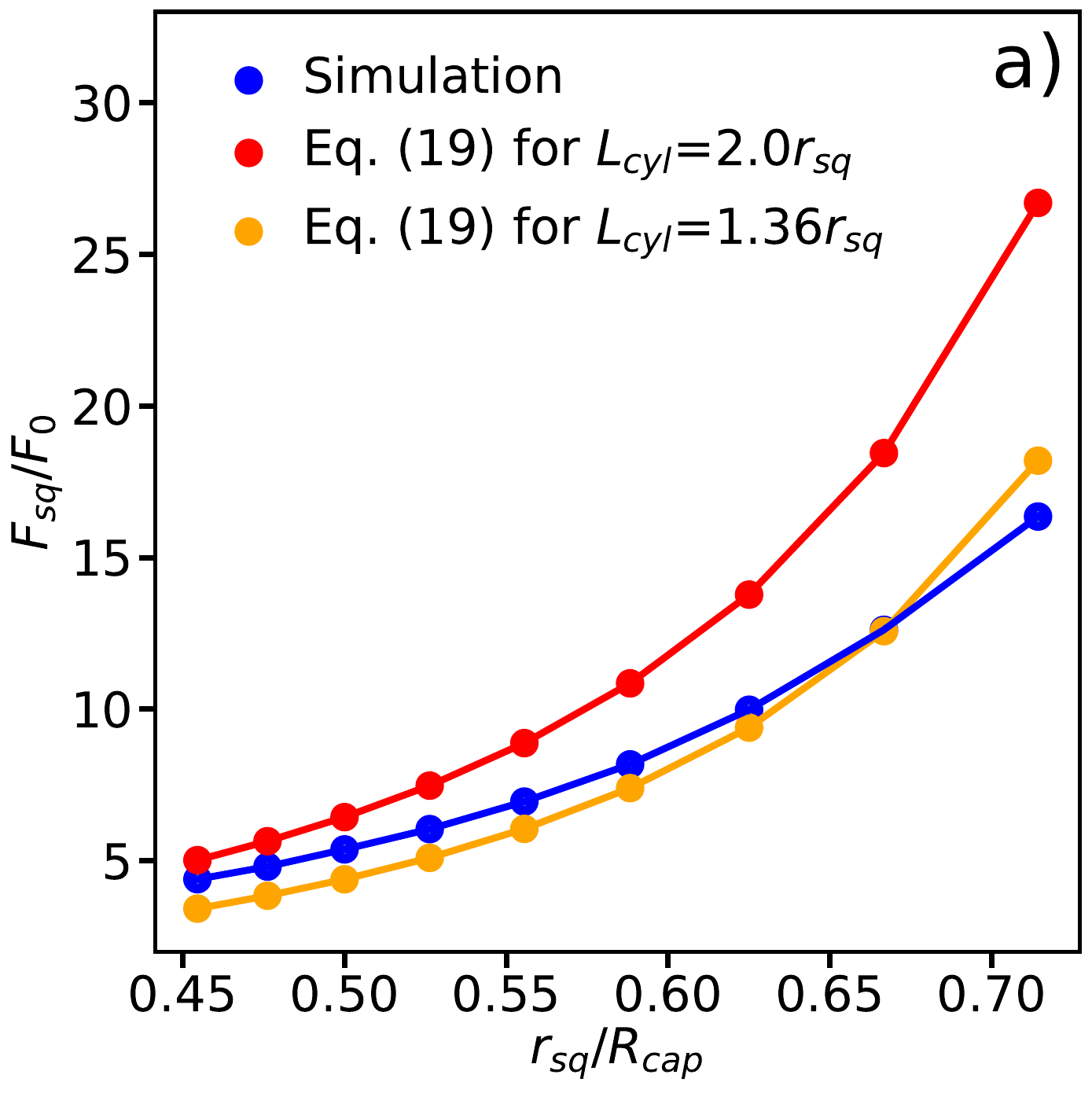}
    \includegraphics[width=0.66\linewidth]{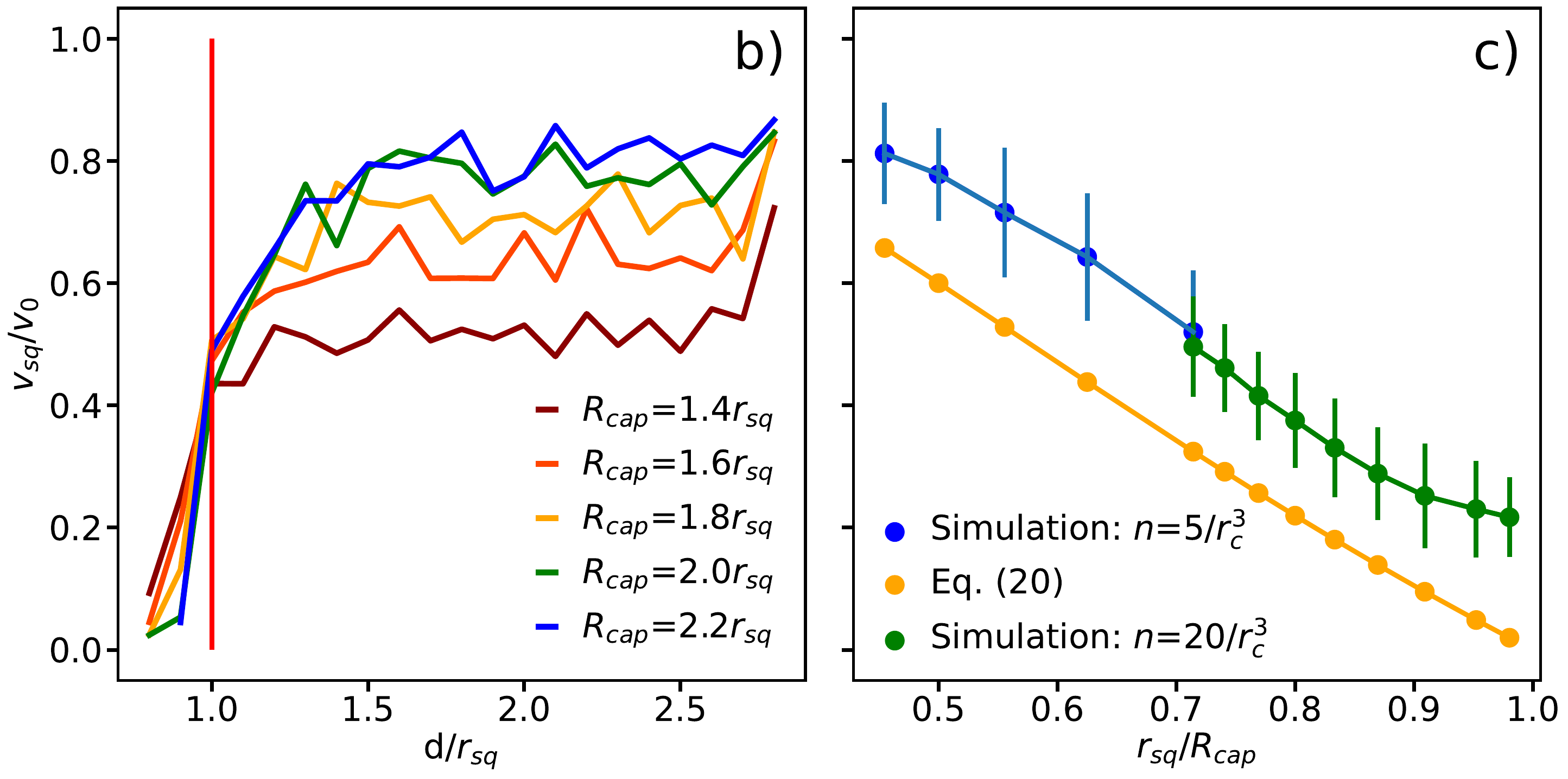}
    \caption{(a) Propulsion force of a spring-fixed spherical squirmer as a function of $r_{sq} / R_{cap}$ in a closed capillary. Simulation results (blue) 
    are compared with the analytical solution (red and orange) from Eq.~(\ref{eq:fiScP_zDeviation}) for two different cylinder lengths and a surface velocity $u_{cyl}=-B_1$. (b-c) Swimming 
    velocity of a free squirmer in a closed tube normalized by $v_0=\frac{2}{3}B_1$. (b) $v_{sq}$ as a function of the distance $d$ from the squirmer's 
    center of mass to the closed end of the tube. Different colors represent different capillary radii, changing the confinement. The vertical red line 
    indicates the distance at which the squirmer touches the wall. (c) Squirmer velocity away from the tube ends (averaged within the region $0.2r_{sq} \leq z \leq r_{sq}$) in comparison 
    with the analytical model from Eq.~(\ref{eq:frSqcP_velocity}) for $u_{cyl} = -v_0$. Simulation results are presented for 
    different fluid resolutions with $n = 5/r_c^3$ (blue) and $n = 20/r_c^3$ (green).}
    \label{fig:fiScP-x0AB1}
\end{figure*}

\subsection{Simulation of a squirmer in a periodic capillary}
\label{subsection:simulating_a_micro_swimmer_in_open_pipe}

A spring-fixed squirmer, whose center of mass is subject to a harmonic spring force $F=-kz$, starts swimming and comes to a halt 
when its propulsion force is equal to the spring force. Figure \ref{fig:fiSqoP_sim-radialFlowProfile} compares the fluid velocity profile in the gap obtained 
from a simulation with the corresponding analytical prediction in Eq.~(\ref{eq:ugap}). The velocity is normalized by $v_0 = \frac{2}{3}B_1$, which is the swimming 
velocity of a squirmer without confinement (or when $r_{sq} / R_{cap} \to 0$). 
The qualitative characteristics of $u_{gap}(r)$ are in a good agreement between the simulation and theory.
Under the assumption that the cylinder length is equal to the squirmer diameter (i.e. $L_{cyl}=2r_{sq}$), flow velocity in the gap is slightly faster for the analytical model than in the simulation.
When we do not fix $L_{cyl}$, but use it as a fitting parameter, the best fit between the theory and simulation is found for 
$L_{cyl}=1.41r_{sq}$. This value is somewhat smaller than the diameter of the squirmer, consistent with our expectations.

The extension of the spring by the spring-pinned squirmer allows the quantification of the propulsion force $F_{sq}$. Figure \ref{fig:fiSoP_sim-x0AR} (a) presents the dependence of $F_{sq}$ on the confinement $D$. As expected 
from the analytical model in Eq.~(\ref{eq:F_sigma_open_pipe}), $F_{sq}$ increases with increasing $D$. Comparison of the simulation results 
with the analytical solution shows a good agreement for a fitted cylinder length of $L_{cyl} = 1.22r_{sq}$, while the choice of $L_{cyl}=2r_{sq}$ in
Eq.~(\ref{eq:F_sigma_open_pipe}) results in the overprediction of the propulsion force measured in simulations. Figure \ref{fig:fiSoP_sim-x0AR} (b) 
shows an increase in the propulsion force of the squirmer as a function of the capillary length $L_{cap}$, which can be fitted well by the analytical model 
with $L_{cyl}=1.25r_{sq}$. The increase in $F_{sq}$ with increasing $L_{cap}$ is likely due to an increased friction for fluid flow in longer 
capillaries, which leads to slower flow velocities within the tube and more efficient swimmer propulsion. 

Figure \ref{fig:fiSoP_sim-x0AR} (c-d) shows the swimming velocity of a free squirmer as a function of $D$ and $L_{cap}$. As expected, all simulated 
velocities of the squirmer are smaller than $v_0=\frac{2}{3}B_1$ for a squirmer in an unbounded fluid. The description of the simulation data 
by the analytical model from Eq.~(\ref{eq:v_anal_open_pipe}) is less accurate here. We choose $u_{cyl}$ such that the swimming velocity of the theoretical model matches the bulk velocity of a squirmer for $D\to 0$. At high confinement, 
there seems to be a qualitative disagreement between simulations and the analytical model [see Fig.~\ref{fig:fiSoP_sim-x0AR} (c)], which is due 
to an insufficient fluid resolution in simulations when the gap between the squirmer and the tube becomes very narrow. We have performed simulations 
with a four times larger fluid density ($n = 20/r_c^3$), which show that the nominal resolution with $n=5/r_c^3$ is sufficient only up to confinements of 
$D \lesssim 0.75$. Simulations with the finer resolution do reproduce an increase in 
$v_{sq}$ at large confinements, in qualitative agreement with the analytical model. Furthermore, our swimming velocity converges for increasing capillary length to that in Ref.~\cite{Zhu_LRN_2013}, where the swimming velocity of $v_{sq} = 0.8 v_0$ was found for the confinement of $D=0.5$.

We have also performed a few simulations using a spheroidal squirmer for intermediate confinements (see Fig.~\ref{fig:fiSoP_sim_spheroidal}). The qualitative trends 
of spheroidal squirmer motion as a function of confinement are the same as for the spherically-shaped squirmer, in agreement with the analytical model. 
Further simulations showed that the swimming velocity of the spheroidal squirmer is larger than for the spherical case, which is in agreement with theoretical predictions \cite{Theers_MSM_2016}.

\begin{figure}[!th]
    \centering
    \includegraphics[width=\linewidth]{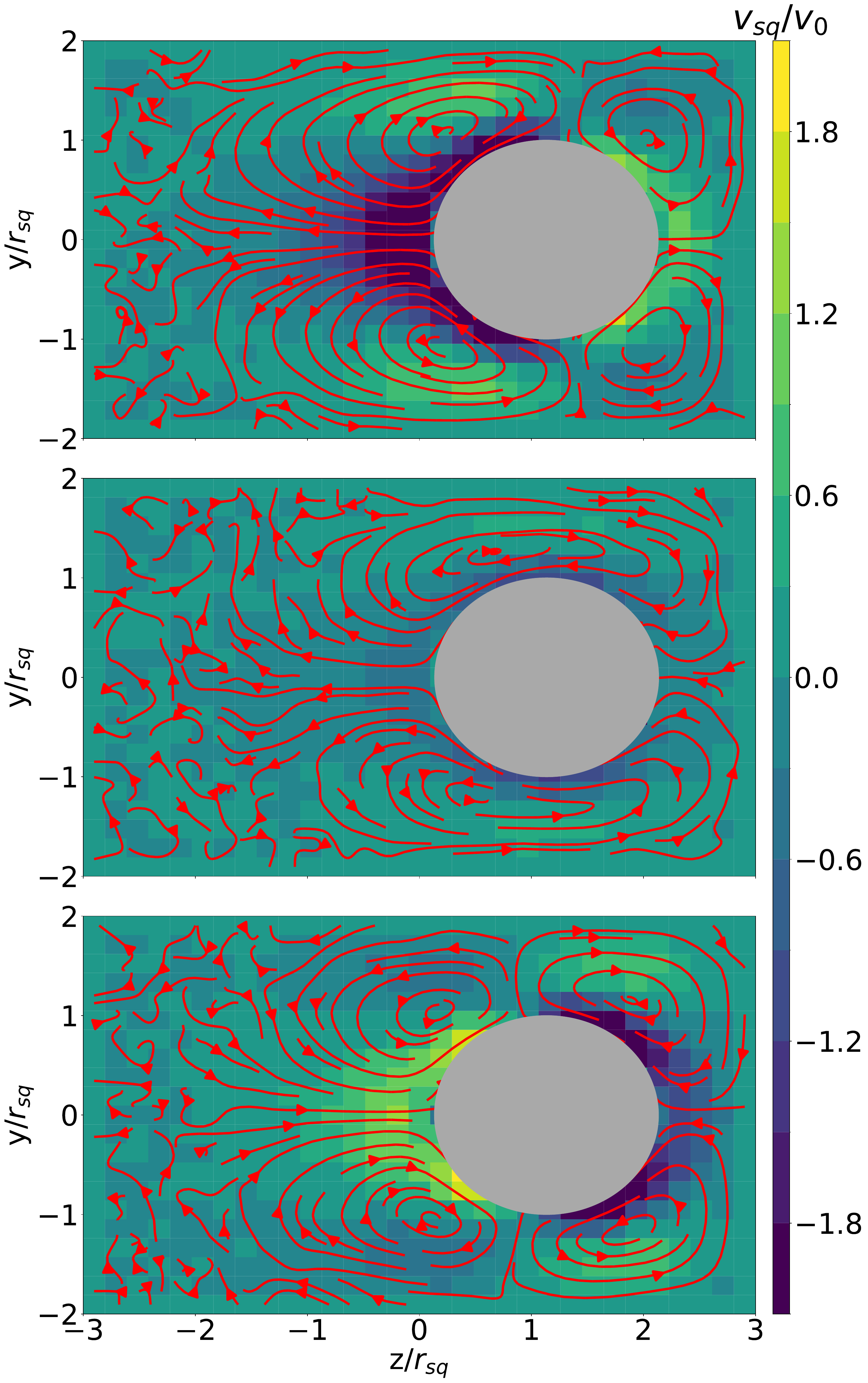}
    \caption{Flow field around a spherical squirmer (grey) within a closed capillary with a length of $L_{cap} = 6 r_{sq}$ and a radius of $R_{cap}=2 r_{sq}$. 
    The squirmer is tethered by a spring at the anchoring point with $z_{anchor} = r_{sq}$. The colors code the flow velocity in the $z$ 
    direction normalized by $v_0=\frac{2}{3}B_1$. The red arrows represent flow lines within the y-z-plane. From the top to the bottom, 
    the figures correspond to pusher ($\beta=-5$), neutral ($\beta=0$), and puller ($\beta=5$) squirmers.}
    \label{fig:flowfieldagainstb}
\end{figure}

\subsection{Simulation of a squirmer in a closed capillary}
\label{subsection:simulating_a_micro_swimmer_in_an_closed_pipe}

Figure \ref{fig:fiScP-x0AB1} (a) shows the propulsion force of a spherical spring-pinned squirmer in a closed tube as a function of confinement. The 
generated force increases with increasing $r_{sq} / R_{cap}$ and the correspondence between simulations and the analytical model in 
Eq.~(\ref{eq:fiScP_zDeviation}) is good for $L_{cyl} = 1.35 r_{sq}$. Figure \ref{fig:fiScP-x0AB1} (b) presents the swimming velocity 
of a neutral squirmer as a function of the distance between its center of mass and the wall at the positive end of the tube for different 
confinements. The swimming velocity becomes slower in more confined systems. Interestingly, the effect of the wall at the positive end of 
the tube on swimming velocity becomes relevant only for a distance $d \lesssim 1.5 r_{sq}$, corresponding to a distance of half a radius between the 
wall and the squirmer surface. This indicates that the flow field generated by the squirmer is local and does not extend beyond 
the distance of $0.5r_{sq}-0.8 r_{sq}$ away from the squirmer surface. Note that $d$ can be slightly lower than $r_{sq}$, because the squirmer slightly deforms when it directly interacts 
with the wall. Nevertheless, away from the wall the squirmer shape remains spherical.

Figure \ref{fig:fiScP-x0AB1} (c) compares the squirmer velocity for $d > r_{sq}$ with the results of the analytical model from Eq.~(\ref{eq:frSqcP_velocity}).
The swimming velocity reduces with increasing confinement for both simulations and the analytical model, but simulations display overall larger velocities. Note that for confinements $D>0.75$, we employ simulations with an increased fluid resolution ($n=20/r_c^3$). 
A few simulations with a spheroidal squirmer shape show qualitatively similar behavior of the propulsion force and the swimming velocity as for the case of spherical squirmer. 
These results further support the validity of the proposed analytical model of squirmer propulsion under confinement.

\begin{figure}[!th]
    \centering
    \includegraphics[width=0.8\linewidth]{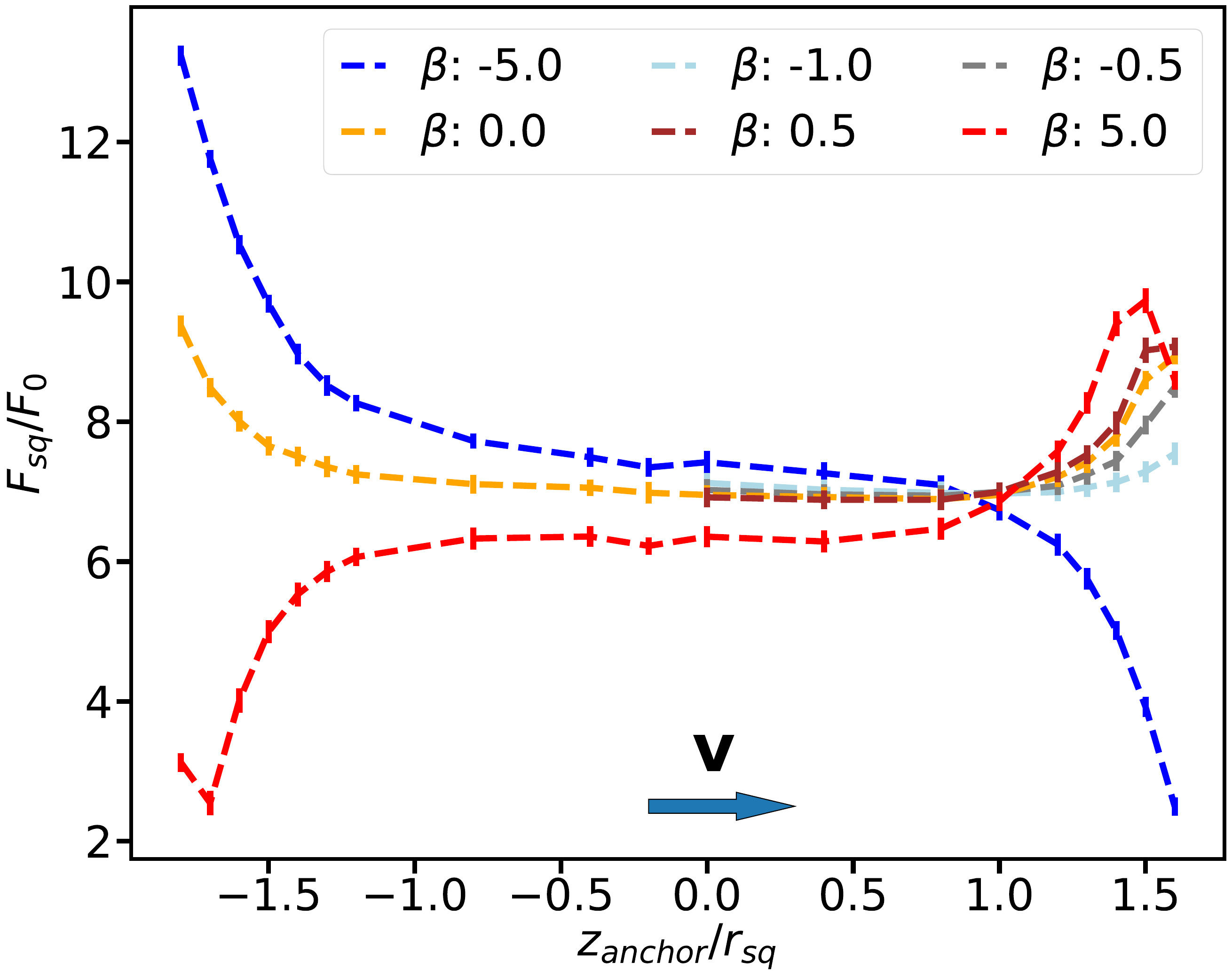}
    \caption{Propulsion force of a spherical spring-fixed squirmer as a function of its anchoring position $z_{anchor}$ for different swimming modes 
    in a closed capillary with $L_{cap} = 6 r_{sq}$ and $R_{cap}=2 r_{sq}$. $z_{anchor}=0$ corresponds to the anchoring point in the middle of the tube length. 
    The swimming orientation of the squirmer is always in the 
    positive $z$ direction, while the anchoring point also includes negative $z$ values. The colors represent various swimming modes with 
    different $\beta$ values.}
    \label{fig:fiScPx0AB2b}
\end{figure}

\subsection{Effect of the swimming mode on squirmer behavior}
\label{subsection:The_affect_of_the_swimming_mode_on_the_propulsion_force}

All simulations described in Sections~\ref{subsection:simulating_a_micro_swimmer_in_open_pipe} and \ref{subsection:simulating_a_micro_swimmer_in_an_closed_pipe} are for a neutral squirmer with $\beta=0$ (or $B_2=0$), for which the slip velocity field is symmetric with respect to the 
squirmer's center of mass. Other swimming modes are obtained by varying the parameter $\beta$, which generates pusher ($\beta < 0$) 
and puller ($\beta > 0$) modes, where the thrust is generated predominantly in the front and in the back of the squirmer, respectively. Note that $\beta \neq 0$ 
affects local flow field around the squirmer, but does not change its swimming velocity in an unbounded fluid. However, $\beta$ strongly affects 
the interaction of the squirmer with walls and other swimmers \cite{Gotze_MSH_2010,Zoettl_NDM_2012,Berke_HAS_2008}. 
For example, a recent simulation study has shown that the swimmer navigation through a regular lattice of solid spheres depends on the swimming mode of microswimmers \cite{Chamolly_APL_2017}. 
It is important to note that the proposed analytical model does not include the effect of $\beta \neq 0$.

We consider the case of a closed capillary, because the propulsion velocity of squirmers in a capillary with periodic BCs has only a weak dependence on the swimming mode, while 
squirmer interaction with the closed ends of the capillary is very different for pusher, neutral, and puller squirmers. Figure \ref{fig:flowfieldagainstb} 
shows flow fields around the squirmer spring-anchored at $z_{anchor} = r_{sq}$ for different swimming modes. The flow fields are qualitatively different, and are significantly affected by confining walls. Figure \ref{fig:fiScPx0AB2b} presents propulsion forces of different squirmers 
anchored at various positions $z_{anchor}$ along the tube axis. When the squirmer is far enough ($L_{cap} /2 - |z_{anchor}| \gtrsim 1.5 r_{sq}$) from the closed ends 
of the capillary, its propulsion force is independent of the anchoring position. The propulsion force of the neutral squirmer increases as the squirmer approaches the front end of the tube. The 
increase in ${\bf F}_{sq}$ for the neutral squirmer appears to be similar for the both tube ends, indicating that the squirmer interaction 
with the walls is independent of whether it swims away from or toward one of the closed ends. This is likely due to the symmetry of the local flow field 
around the neutral squirmer along its propulsion direction. 
When the squirmer is fixed at the center of the tube, ${\bf F}_{sq}$ is slightly larger (smaller) for the pusher (puller) in comparison to the neutral squirmer.
The puller generates larger propulsion forces than the neutral squirmer when approaching 
the closed end in swimming direction. However, when the squirmer swims away from the capillary end, the puller 
is weaker than the neutral squirmer. Swimmer interaction with the closed ends is opposite for pushers in comparison with pullers. 
The pusher generates larger propulsion forces when it moves away from the tube end, and smaller forces when it moves toward the end in comparison to those of 
the neutral squirmer. This is not entirely surprising, because the asymmetry of the generated flow field is directed toward the front for pullers
and toward the back for pushers. We have also verified that spheroidally-shaped squirmers exhibit qualitatively similar behavior in a closed capillary for different swimming modes.

\section{Discussion and conclusions}
\label{discussion}

We have derived an analytical model for squirmer propulsion under confinement, and validated it using simulations of spherical and 
spheroidal squirmers moving inside a capillary. 
Both the analytical model and simulations demonstrate that the locomotion of squirmers is possible in cylindrical capillaries even under very strong confinement. For squirmers,
the propulsion force increases with increasing confinement, and diverges as $r_{cyl}/ R_{cap} \rightarrow 1$, according to predictions of the analytical 
model. This is in qualitative agreement with theoretical calculations for a swimmer near a wall \cite{Ishikawa_SCG_2019}. Despite the fact that the propulsion 
force may become very large under strong confinements, the swimming velocity can never exceed the velocity of a swimmer in an unbound fluid for 
$r_{sq}/R_{cap} \rightarrow 0$. This indicates that the fluid resistance generally grows faster than the propulsion force with increasing 
confinement.

The modeled swimmers generate their propulsion through the prescribed surface velocity, which results in large propulsion forces under strong confinements,
independently of the magnitude of fluid resistance. In the theoretical model, these swimmers can generate an infinite power, which is clearly not the case 
for real biological or synthetic microswimmers, whose ability to generate propulsion has an upper limit. This indicates that real microswimmers 
would likely move slower under strong confinement in comparison to the predictions by the squirmer model, or may even stop moving for a fixed propulsion force when $r_{sq}/R_{cap} \rightarrow 1$. 
Furthermore, details of the local flow field generated by a microswimmer are important for its propulsion through confinements and crowded environments.
Since biological swimmers generally propel due to the motion of attached appendages such as flagella and cilia, steric interactions between the appendages 
and surrounding walls can also affect the navigation of microswimmers through crowded environments. To accurately capture the complex interactions 
between microswimmers and their environment, more realistic models of biological swimmers with explicit appendages are required.

An important result of our investigation is that the generation of propulsion forces by squirmers under confinement is very localized, and can be thought of as 'pushing forward' using the walls, which occurs due to the interaction of local flow field generated by the swimmer with the walls. 
This is well supported 
by a short distance beyond which the generated flow field vanishes, see Fig.~\ref{fig:flowfieldagainstb}, and by the fact that the presence of closed tube ends 
is not important if the swimmer is more than its radius away from them. Therefore, microswimmers employ predominantly the closest confining walls around them to propel forward. 
Also, this means that specific geometric details of a confining environment are increasingly important close to the swimmer, and become irrelevant further away from it.
Furthermore, the locality of force generation is relevant for the understanding of tether pulling from fluid membrane vesicles by enclosed microswimmers
\cite{Vutukuri_APV_2020,Takatori_ACF_2020,Le_Nagard_EBV_2022}. For instance, we conclude that a squirmer-like swimmer cannot pull long tethers,
because when the squirmer becomes fully and closely surrounded by the membrane after tether initiation, the generated force cannot propagate anymore toward the vesicle through the 
tether in order to further extend it. Furthermore, the force on the tether end is compensated by an opposing force on the tether walls. In a recent experiment of {\it Bacillus subtilis} bacteria in fluid vesicles, relatively short tethers filled by a train-like arrangement of $2-3$ bacteria are 
formed \cite{Takatori_ACF_2020}. It is likely due to sequential entering of bacteria, such that the first initiates a tether of approximately its own length, then a second 
bacterium extends it by another bacterium length, etc. In another experiment \cite{Le_Nagard_EBV_2022}, {\it E. coli} bacteria were able to pull 
relatively long tethers due to another mechanism, where {\it E. coli} helical flagella form a single bundle, which is tightly wrapped by the membrane after 
tether formation, so that the membrane-wrapped bundle serves as a propeller to move forward and extend the tether. 

Our simulations show that strong confinements may require a fine resolution to accurately capture fluid flow between the swimmer and the wall, 
which is associated with large computational costs. Furthermore, fluid compressibility may play an important role for systems with strong confinements, 
as large pressure gradients develop. The DPD method has a limited control over the fluid compressibility, which can be improved by using the 
smoothed dissipative particle dynamics method \cite{Espanol_SDPD_2003,Mueller_SDPD_2015}, where the equation of state can be prescribed explicitly. 

\section*{Appendix A: Spheroidal squirmer model}
\label{subsection:the_spheroidal_squirmer_model}

For the derivation of analytical expressions, it is convenient to work in modified prolate spheroidal coordinates \cite{Poehnl_ASS_2020} $\zeta \in [1, -1]$, 
$\tau \in [1, \infty]$, and $\phi \in [0, 2\pi]$, whose relation to the cartesian coordinates is given by 
\begin{align}
    \tau &= \frac{1}{2c}\Big( \sqrt{x^2+y^2+(z+c)^2}+\sqrt{x^2+y^2+(z-c)^2}\Big), \nonumber  \\
    \zeta &= \frac{1}{2c}\Big( \sqrt{x^2+y^2+(z+c)^2}-\sqrt{x^2+y^2+(z-c)^2}\Big),  \nonumber \\
    \phi &= \arctan(\frac{y}{x}),
\end{align}
where $c \equiv \sqrt{b_z^2-b_x^2}$ is the geometric constant which controls spheroid eccentricity $e = c/b_z$. Normal (${\bf n} = {\bf e}_\tau$) and 
tangential (${\bf t} = -{\bf e}_\zeta$) vectors at the spheroid surface are given by
\begin{align}
    {\bf e}_\tau &= \Big( \frac{2\tau\sqrt{1-\zeta^2}}{\sqrt{\tau^2-1}}{\bf e}_x + \zeta{\bf e}_z\Big) \frac{\sqrt{\tau^2-1}}{\sqrt{\tau^2-\zeta^2}},    \nonumber \\
    {\bf e}_\zeta &= \Big(\frac{2\tau\sqrt{\tau^2-1}}{\sqrt{1-\zeta^2}}{\bf e}_x + \tau{\bf e}_z\Big) \frac{\sqrt{1-\zeta^2}}{\sqrt{\tau^2-\zeta^2}}.
\end{align}
The surface velocity of a spheroidal squirmer is then \cite{Theers_MSM_2016}
\begin{equation}
\label{eq:surface_velocity_spheroidal}
    {\bf u}_{surf} = -B_1(1+\beta\zeta)({\bf t}\cdot{\bf e}_z){\bf t},
\end{equation}
where $\beta = B_2/B_1$ and ${\bf e}_z = (0, 0, 1)^T$.

\section*{Appendix B: Expressions for the parameters of the analytical model for cylindrical microswimmers}
\label{sec:materials_and_methods}
Parameters for the case of a spring-fixed swimmer in a periodic capillary are given by
\begin{align}
\label{eq:fisqop_constants}
    c_2 &= -\frac{\Delta p}{4\eta L_{cyl}}{R_{cap}}^2-c_1\ln(R_{cap}), \nonumber \\
    c_1 &= \frac{u_{cyl}+\frac{\Delta p{R_{cap}}^2}{4\eta L_{cyl}}(1-D^2)}{\ln(D)}, \nonumber \\
    \Delta p &= -\frac{4u_{cyl}\eta}{{R_{cap}}^2}\frac{1-D^2+2D^2\ln(D)}{\frac{1}{L_{cyl}}((1-D^4)\ln(D)+(1-D^2)^2)+\frac{\ln(D)}{L_{cap}-L_{cyl}}}.
\end{align}

For a free cylindrical swimmer in a periodic capillary, the parameters are
\begin{align}
    c_2 &= -\frac{\Delta p}{4\eta L_{cyl}}R_{cap}^2, \nonumber \\ 
    c_1 &= 0, \nonumber \\
    \Delta p &= -\frac{u_{cyl}8\eta}{R_{cap}^2}\frac{D^2}{(1-D^4)/L_{cyl}+1/(L_{cap}-L_{cyl})}.
\end{align}

Paramters for the case of a spring-pinned swimmer in a closed capillary are given by
\begin{align}
    c_2 &= -\frac{\Delta p}{4\eta L_{cyl}}R_{cap}^2-c_1\ln(R_{cap}), \nonumber \\
    c_1 &= \frac{u_{cyl}+(\Delta p R_{cap}^2)/(4\eta L_{cyl})\cdot(1-D^2)}{\ln(D)} = \frac{u_{cyl}}{1+\ln(D)(1+D^2)/(1-D^2)}, \nonumber \\
    \Delta p &= -\frac{u_{cyl}4\eta L_{cyl}}{R_{cap}^2}\frac{1-D^2+2D^2\ln(D)}{(1-D^2)^2+\ln(D)(1-D^4)}.
\end{align}

For a free cylindrical swimmer in a closed capillary, the parameters are
\begin{align}
    c_2 &= -\frac{\Delta p R_{cap}^2}{4\eta L_{cyl}} = 2u_{cyl}\frac{D^2}{D^4-1}, \nonumber \\
    c_1 &= 0, \nonumber \\
    \Delta p &= \frac{u_{cyl}8\eta L_{cyl}}{R_{cap}^2}\frac{D^2}{D^4-1}.
\end{align}

\section*{Author Contributions}

G.G. and D.A.F. conceived the research project. F.O. and D.A.F. developed the analytical model. F.O. performed the simulations and analysed the obtained data. 
All authors participated in the discussions and writing of the manuscript.

\section*{Conflicts of interest}

There are no conflicts to declare.

\section*{Acknowledgements}

Support by the Deutsche Forschungsgemeinschaft (DFG) within the Priority Programme "Physics of Parasitism" (SPP 2332) is gratefully acknowledged.
The authors gratefully appreciate computing time on the supercomputer JURECA \cite{jureca} at Forschungszentrum J{\"u}lich under grant no. actsys.



%

\end{document}